%% file: Report19.tex
\documentclass[final]{IEEEtran}
\usepackage{amsthm,amssymb,graphicx,multirow,amsmath,color,amsfonts}
\usepackage[update,prepend]{epstopdf}
\usepackage[noadjust]{cite}
\usepackage[latin1, utf8]{inputenc}
\usepackage{tikz}
\usepackage{bbm} 
\usepackage{pdfpages}
\usepackage{tabulary}
\usepackage{multirow}
\usepackage{comment}
\usepackage{subfigure}
\usepackage{float}
\usepackage[titlenumbered,ruled]{algorithm2e}
\usepackage[noend]{algpseudocode}
\DeclareUnicodeCharacter{0177}{\^y}

\include{notation}

\allowdisplaybreaks 
\begin{document}
\title{Velocity-Aware Statistical Analysis of Peak AoI for Ground and Aerial Users
}
\author{
	Yujie Qin, Mustafa A. Kishk, {\em Member, IEEE}, and Mohamed-Slim Alouini, {\em Fellow, IEEE}
	\thanks{Yujie Qin is with the School of Information and Communication Engineering, University of Electronic Science and Technology of China (UESTC), Chengdu 610054, China.  Mustafa Kishk is with the Department of Electronic Engineering, Maynooth University, Maynooth, W23 F2H6, Ireland. Mohamed-Slim Alouini is with Computer, Electrical and Mathematical Sciences and Engineering (CEMSE) Division, King Abdullah University of Science and Technology (KAUST), Thuwal, 23955-6900, Saudi Arabia. (e-mail: yujie.qin@uestc.edu.cn; mustafa.kishk@mu.ie; slim.alouini@kaust.edu.sa).} 
}
\date{\today}
\maketitle

\begin{abstract}
	In this paper, we present a framework to analyze the impact of user velocity on the distribution of the peak age-of-information (PAoI) for both ground and aerial users by using the dominant interferer-based approximation. We first approximate the SINR meta distribution for the uplink transmission using the distances between the serving base station (BS) and each of the user of interest and the dominant interfering user,  which is the interferer that provides the strongest average received power at the tagged BS. We then analyze the spatio-temporal correlation coefficient of the conditional success probability by studying the correlation between the aforementioned two distances. Finally, we choose PAoI as a performance metric to showcase how spatio-temporal correlation or user velocity affect system performance.
	Our results reveal that ground users exhibit higher spatio-temporal correlations compared to aerial users, resulting in a more pronounced impact of velocity on system performance, such as joint probability of the conditional success probability and distribution of PAoI.
	Furthermore, our work demonstrates that the dominant interferer-based approximation for the SINR meta distribution delivers good matching performance in complex scenarios,  such as Nakagami-m fading model, and it can also be effectively utilized in computing spatio-temporal correlation, as this approximation is derived from the distances to the serving BS and the dominant interferer. 
\end{abstract}

\section{Introduction}

With the development of advanced wireless communication and sensing techniques, Internet of Things (IoT) devices are deployed widely to facilitate several real-time applications ranging from localization \cite{li2020toward} and remote monitoring to image collection and environmental sensing \cite{10104086,shit2018location}. 
The consequential deployment of such IoT devices underscores the paramount importance of establishing low latency and reliable communication channels, particularly in scenarios where time-sensitive data exchange is imperative. 
A primary example lies in the domain of intelligent transportation systems (ITS), which encompass various modes of transportation, including roads, air, and water transportation, with a primary focus on enhancing the safety and efficiency of these networks \cite{7331645}. Within ITS, IoT devices emerge as indispensable contributors to road safety, adeptly gathering and transmitting real-time data pertaining to traffic patterns, congestion levels, and road conditions.
This real-time information not only empowers drivers to make dynamic decisions but also actively reduces the likelihood of accidents, simultaneously laying the foundation for the eventual integration of autonomous and semi-autonomous vehicles \cite{9114879}.

To this objective, collecting accurate and up-to-date data, the information age, which characterizes the freshness of the data, draws great attention.  In contrast to conventional metrics such as delay, the concept of the `age of information' (AoI) emerges as a more nuanced and insightful perspective \cite{kaul2012real,alberts1997information}. AoI is defined as the duration elapsed since the last successfully received update packet at the monitor was generated at the source, effectively encapsulating the timelines of updates \cite{emara2020spatiotemporal}. As new updates flow in, the AoI incrementally grows until the next update arrives, essentially measuring the time span between two successfully transmitted updates. Consequently, the efficient collection of data from IoT devices and the reduction of AoI at the receiver's end become pivotal elements in preserving the functionality and ensuring a high quality of service for the network.

Generally, the analysis of AoI of IoT devices confines to static devices or those fixed at specific locations \cite{9042825}. However, as mentioned in previous text, devices can be deployed on moving vehicles, continuously communicating with nearby base stations (BSs) as they travel through different areas. In these scenarios, the analysis of static devices show limitations and cannot fully reveal the system insights of moving devices.
 For instance, if a moving IoT device transmits an update to a BS at two different times (assuming no handover occurs), these two transmission times are not entirely independent, as the time is influenced by the distance to the serving BS, which demonstrates spatio-temporal correlation.
Delving into the analysis of this spatio-temporal dependence, particularly in relation to the velocity and arrival process of updates, introduces the need for a comprehensive examination of observations made at distinct spatio-temporal locations. This presents a more intricate challenge compared to the conventional spatial average AoI analysis. For instance, analyzing the AoI of a moving device requires the computation of the joint distribution of the transmission time at different time instants.

Noticing the ubiquity of stochastic geometry as a prevalent tool for characterizing system performance, its efficacy is underscored by its precision in modeling the spatial distribution of network elements, such as locations of BSs, users,  and randomness of channel fading. This tool also enables the derivation of closed-form expressions in special cases \cite{haenggi2009stochastic,elsawy2016modeling,elsawy2013stochastic,stoyan2013stochastic}.
{\color{blue}Motivated by the fact that AoI-related analysis is mainly confined to static devices and lacks spatio-temporal correlation investigation, we use tools from stochastic geometry to analyze the impact of IoT device velocity on AoI. }This analysis encompasses both aerial users, such as unmanned aerial vehicles (UAVs), and ground users, including cars and trains.

\subsection{Related Work}
Literature related to this work can be categorized into: (i) performance analysis of velocity-aware networks, (ii) large-scale-related analysis for AoI or PAoI, and (iii) stochastic geometry-based analysis for spatio-temporal correlation. A brief discussion on related works in each of these categories is discussed in the following lines.

{\em Performance analysis of velocity-aware networks.} 
In velocity-aware networks, stochastic geometry-based analysis often revolves around the critical area of handover management. In \cite{7827020}, the authors examined handover management in two-tier cellular networks. This investigation considered the modeling of BS locations using two independent Poisson point processes (PPPs) and accounted for the velocity of users equipment (UEs). The analysis of cross-tier handover in small cell networks was a key focus in \cite{7248855}, while \cite{8555918} delved into the handover probability in dense cellular networks. The study of RF/VLC hybrid networks was explored in \cite{9445651}, addressing the unique challenges and opportunities presented by such hybrid configurations. For cellular-enabled UAV networks, \cite{9838749} conducted a time-varying SNR analysis, and \cite{9003219} delved into the handover probability of UAV-assisted networks, recognizing the increasing role of UAVs in modern communication systems. The velocity of users can introduce Doppler shifts, impacting system performance. Authors in \cite{9394765} analyzed the performance effects associated with velocity-induced Doppler shifts. Additionally, \cite{9148880} proposed a study examining Doppler shift-related considerations in communication links between satellites and terrestrial users.

{\em Large-scale-related analysis for AoI or PAoI.}  
AoI is a critical metric that quantifies the freshness of information in IoT devices. In \cite{yates2017status}, a general introduction and survey of AoI was presented, offering a broad perspective on the concept's applications in IoT networks. The intersection of AoI and stochastic geometry was briefly introduced in \cite{8930830}, which explored the relationship between AoI and peak AoI (PAoI). The first application of stochastic geometry to AoI can be found in \cite{yang2020optimizing}, where a Poisson bipolar network was considered, emphasizing the impact of network geometry on AoI.
Different transmission policies, such as first-come-first-serve and last-come-last-serve, were investigated in \cite{yang2021understanding}, shedding light on the influence of policy choices on AoI.
Queuing disciplines, both non-preemptive and preemptive, were employed in \cite{9360520} to compute PAoI in a large-scale network. The study incorporated Poisson bipolar distribution for source-destination pairs' locations.
Device-to-device communication-based AoI analysis was presented in \cite{9467495}, where the network throughput and mean AoI for a cellular-based IoT network were computed.
IoT devices' time- and event-triggered traffic-based PAoI analysis was conducted in \cite{emara2020spatiotemporal} and \cite{emara2020spatiotemporal1}, with a focus on networks where IoT device and BS locations were modeled using two independent PPPs. In \cite{10266697}, a UAV-assisted IoT network was analyzed, highlighting the role of UAVs in collecting data from correlated and uncorrelated devices.
Finally, reinforcement learning frameworks were explored for stochastic geometry-related AoI analysis in \cite{9085402}, and a deep reinforcement learning framework was proposed in \cite{9013924}. These frameworks offer new avenues for optimizing and managing AoI in complex network environments.

{\em Stochastic geometry-based analysis for spatio-temporal correlation.} 
A tutorial on the stochastic geometry-based analysis of spatial-temporal performance in wireless communication networks was introduced in \cite{9516701}. This tutorial provided a concise introduction to the correlation effects of shadowing, obstacles, interference, and the spatio-temporal locations of interferers.
In \cite{krishnan2017spatio}, the authors delved into the analysis of the joint coverage probability in a single-tier network where the velocity of UE was considered. They computed the spatio-temporal interference correlation coefficient and exact joint coverage probability at two distinct spatial locations within cellular networks.
Correlations in interference for different times at the same locations and different locations were examined in \cite{5282357} and \cite{schilcher2013does}, particularly in the context of ad hoc networks.
In \cite{haenggi2012diversity}, the author highlighted the significant impact of interference correlation on the diversity of multi-antenna communications, such as single-input-multiple-output (SIMO), and computed the joint SIR distribution for two-antenna links. Spatial-temporal-based analyses for the peak AoI (PAoI) were conducted in \cite{emara2020spatiotemporal}, where the authors investigated the influence of traffic models on interference density and transmit power, and subsequently, on PAoI.
Additional spatial-temporal interference analyses can be found in \cite{tanbourgi2014effect,tanbourgi2014dual,zhou2016performance}.

{\color{blue}Unlike existing stochastic geometry-based analyses that primarily focus on the spatio-temporal correlation of interference, this work analyzes the spatio-temporal correlation of conditional success probability. This approach provides higher-level information on the statistical performance of each link. It offers a more detailed perspective compared to traditional interference metrics. In contrast to static AoI-related research, our work considers UE with velocity, which is practical for real-world implementation, given that IoT devices are deployed on trains, cars, and trams to monitor velocities and locations.} Moreover, unlike velocity-aware studies that predominantly analyze handover issues, our work examines transmission correlation performance, which is crucial in velocity-aware networks.


\subsection{Contribution}
This paper investigates the impact of UE velocity on the distribution of PAoI and derives the joint distribution of the conditional success probability for ground and aerial users, respectively. Differing from existing literature that mainly focuses on analyzing the spatio-temporal correlation of interference, we examine the correlation of device locations and capture the influence of spatio-temporal correlations on the the distribution of the PAoI.

The contributions of this paper include:
\begin{itemize}
	\item We initially employ the dominant interferer-based approximation to express the conditional success probability as a function of the distance to the serving BS and the distance to the nearest interferer for both ground and aerial UEs. Subsequently, we investigate the spatio-temporal correlation of moving UEs by analyzing the spatio-temporal correlation of the aforementioned two distances.
	\item We expand the dominant interferer-based approximation for SINR meta-distribution to aerial-related networks. Noting the challenges of using the Lambert W function to represent the probability of the conditional success probability exceeding $\gamma$ ($\gamma\in[0,1]$, denotes the reliability of the network.), due to the channel fading being modeled by a Gamma distribution, we employ indicator functions instead. The proposed approximation using indicator functions demonstrates a good match in environments with a low line-of-sight (LoS) probability.
	\item We calculate the correlation coefficient of the conditional success probability for ground and aerial UEs, respectively. We illustrate that the conditional success probability of ground UEs exhibits higher correlation than that of aerial UEs due to the variability of LoS/NLoS links in aerial UEs over time. Additionally, while the mean of PAoI remains constant, we demonstrate the impact of UE velocity on the distribution of PAoI.
\end{itemize}

{\color{blue} The structure of this manuscript is as follows: In Section II, we introduce the system model, defining the spatio-correlation and peak Age of Information (AoI) for both moving aerial and ground users. In Section III, we analyze the performance of the moving networks, deriving the equations for distance distributions, approximations of interference, and computing the spatio-correlation coefficient and AoI for both ground and aerial UEs. Finally, the numerical results are presented in Section IV.
}

\section{System Model}
We consider the uplink communication of a single-tier cellular network where the locations of BSs and UEs are modeled by two independent Poisson point processes (PPPs) $\Phi,\Phi_{u}$ with density $\lambda,\lambda_{u}$, respectively. Fixed BSs and UE with mobility are considered, and at time $t$, the locations of UEs form a point process $\Phi_{u}(t) = \{x_u\in\Phi_u(0): x_u+v t\}$. The velocity of users $v$ is fixed but directions of users are random and independent to other users, e.g., $\kappa\sim{\rm Unif}[0,2\pi)$, where $\kappa$ is the movement direction.

We consider  the UEs can either be ground UEs or aerial UEs. In the case of ground UEs, the received power at the BS is
\begin{align}
p_r = p_t H R^{-\alpha},
\end{align}
where $p_t$ is the transmit power of UEs, $H$ denotes the channel fading, which follows an exponential distribution with a mean of unity, $\alpha$ is the path-loss exponent, and $R$ presents the distance between the UE and the serving BS.

{\color{blue}In the case of aerial UEs, the probability of having a LoS transmission is given by the following equation, while the channel fading is modeled using the Nakagami-m distribution}\footnote{ {\color{blue} Nakagami-m is an empirical model fading model characterizes multipath propagation with more than one LoS and NLoS link compared to Rayleigh and Rician models.This makes it particularly suitable for aerial users, as its parameter can be adjusted to represent different fading conditions: for LoS transmission, $m  = 3$, and for NLoS transmission, $m = 1$. }}. Let $R_l$ and $R_n$ be the horizontal distances to the serving BS, the received power of the BS depends on the link which can be LoS or NLoS, is 
\begin{align}
	p_r &= \left\{
	\begin{aligned}
		&p_t G_l \eta_l R_{l}^{-\alpha_l}, \quad {\rm LoS},\\
		&p_t G_n \eta_n R_{n}^{-\alpha_n}, \quad {\rm NLoS},\\
	\end{aligned}
	\right.
\end{align}
in which $G_l$, $\eta_l$, and $\alpha_l$ are the LoS channel fading, which follows a Gamma distribution with shape and scale parameter $(m_l,1/m_l)$, LoS mean additional loss, and LoS path-loss exponent, respectively, and $G_n$, $\eta_n$, and $\alpha_{n}$ are the NLoS channel fading, which follows a Gamma distribution with shape and scale parameter $(m_n,1/m_n)$, NLoS mean additional loss, and NLoS path-loss exponent, respectively. The probability of establishing LoS/NLoS link between the UE and the serving BS, given the horizontal distance is $r$, is given in \cite{al2014optimal} as,
\begin{align}
	P_{ l}(r) & =  \frac{1}{1+a \exp(-b(\frac{180}{\pi}\arctan(\frac{h}{r})-a))},\label{pl_pn}
\end{align}
where $a$ and $b$ are two environment variables which are influenced by the density of obstacles. Consequently, the probability of NLoS link is $P_{ n}(r)=1-P_{  l}(r)$.

\subsection{Spatio-temporal Correlation}
Our objective is to study the uplink communication between the UEs and BSs, hence, we consider that the UE associates with the nearest BS. That is,  the association regions of UEs and BSs form a Poisson Voronoi tessellation \cite{9444343}, where the seeds are the locations of BSs, and the handover happens when UE moves from one cell to another.

	\begin{assumption}[Density and Locations of the Interferernce]
	Assume that BSs split the resource blocks to serve the UE within the Poisson Voronoi cell, hence, only one UE is active at each resource block and the interference comes from other cells \cite{elsawy2017meta}. Besides, we let $\Phi_{i}$ be the locations of the UE that have the same resource block which is equivalent to the locations of the interference, and $\Phi_{i}\in\Phi_{u}$.
\end{assumption}
We randomly select one UE as the typical UE.  Let $\sigma^2$ be the noise power, and the signal-to-interference-plus-noise ratio (SINR) at the BS which the typical UE is associated with is
\begin{align}
	{\rm SINR} &= \frac{p_r}{I+\sigma^2},
\end{align}
where $I$ denotes the aggregated interference. 
Consequently, the conditional success probability of the typical UE is
\begin{align}
	P_s(\theta) = \mathbb{P}({\rm SINR} > \theta|\Phi_i,\Phi).
\end{align}
Now we consider the mobility of UE, while the typical UE moves from $t_0$ to $t_1$, the conditional success probability changes from $P_s(\theta,t_0)$ to $P_s(\theta,t_1)$, and $P_s(\theta,t_0)$ and $P_s(\theta,t_1)$ are correlated due to the spatio-temporal correlation of UE. 

\begin{figure}[ht]
	\centering
	\subfigure[No handover.]{\includegraphics[width = 1\columnwidth]{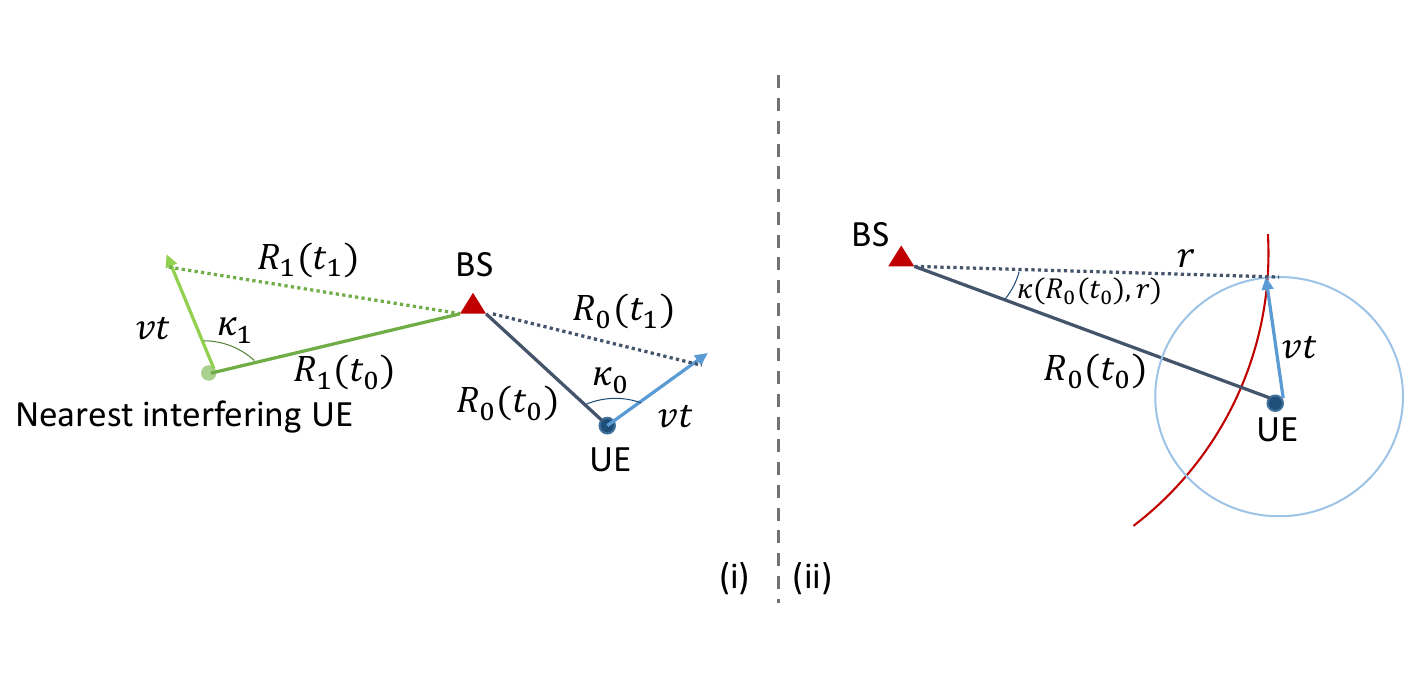}}
	\subfigure[Handover.]{\includegraphics[width = 1\columnwidth]{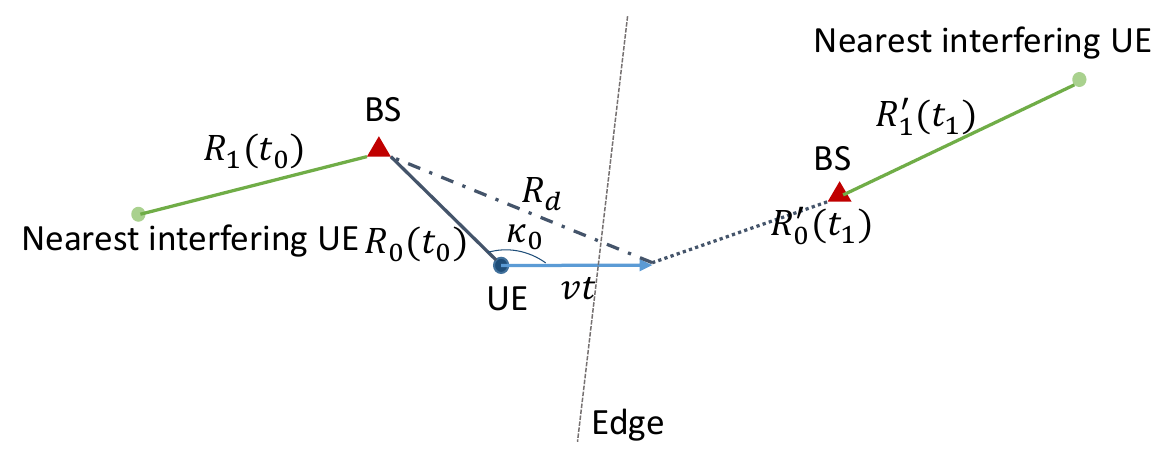}}
	\caption{Illustration of the distances based on UE displacement.  \textbf{(a)} (i) Illustration of the spatial correlation between the distances to the nearest interferer and the serving UE when no handover happens (the BS has the same nearest interferer UE). \textbf{(a)} (ii) Illustration of the distance to the serving BS when no handover happens. \textbf{(b)} Illustration of the spatial correlation between the distances to the nearest interferer and the serving UE when handover happens (BSs have two different nearest interferer UE).}
	\label{fig_sys_handover}
\end{figure}
As shown in Fig. \ref{fig_sys_handover} (a), let $R_0(t_0)$ be the distance from the typical UE to the nearest BS, $v$ and $\kappa_0$ are the velocity and direction of the typical UE at time $t_0$; if no handover, the distance to the nearest BS at time $t_1$ is
\begin{align}
	R_0(t_1) &= \sqrt{(vt)^{2}+R_{0}^{2}(t_0)-2vtR_{0}(t_0)\cos(\kappa_0)},\label{eq_R0R1}
\end{align}
and if handover happens, as shown in Fig. \ref{fig_sys_handover} (b), which means the UE moves to a closer BS, then the distance is
\begin{align}
	R_{0}^{\prime}(t_1) < R_{d} =  \sqrt{(vt)^{2}+R_{0}^{2}(t_0)-2vtR_{0}(t_0)\cos(\kappa_0)},
\end{align}
and  the averaged handover probability is computed in \cite{krishnan2017spatio} as
\begin{align}
	&\mathbb{P}(\mathcal{H}) =\mathbb{E}_{\{\kappa_0,R_0(t_0)\}}\bigg[1-\exp\bigg(-\lambda\bigg(R_{0}^2(t_1)\bigg[\kappa_0\nonumber\\
	&+\sin^{-1}\bigg(\frac{v\sin(\kappa_0)}{R_{0}(t_1)}\bigg)\bigg]-R_{0}^2(t_0)\kappa_0+R_{0}(t_0)v\sin(\kappa_0)\bigg)\bigg)\bigg].\label{eq_ave_handprob}
\end{align}

	We make the following assumption to handle the handover and interference correlation problem.
\begin{assumption}[Handover and Interference Assumption]
	From the perspective of the typical UE, we assume that the locations of interferers are correlated ($\Phi_{i}(t_1) = \{x\in\Phi_i(t_0): x+v (t_1-t_0)\}$) if no handover happens \cite{krishnan2017spatio}, and the interferers are uncorrelated ($\Phi_{i}(t_1) \in\Phi_{u}$) if the handover happens due to the random allocation of the resource blocks \cite{7827020}.
\end{assumption}

\begin{figure}[ht]
	\centering
	\includegraphics[width = 1\columnwidth]{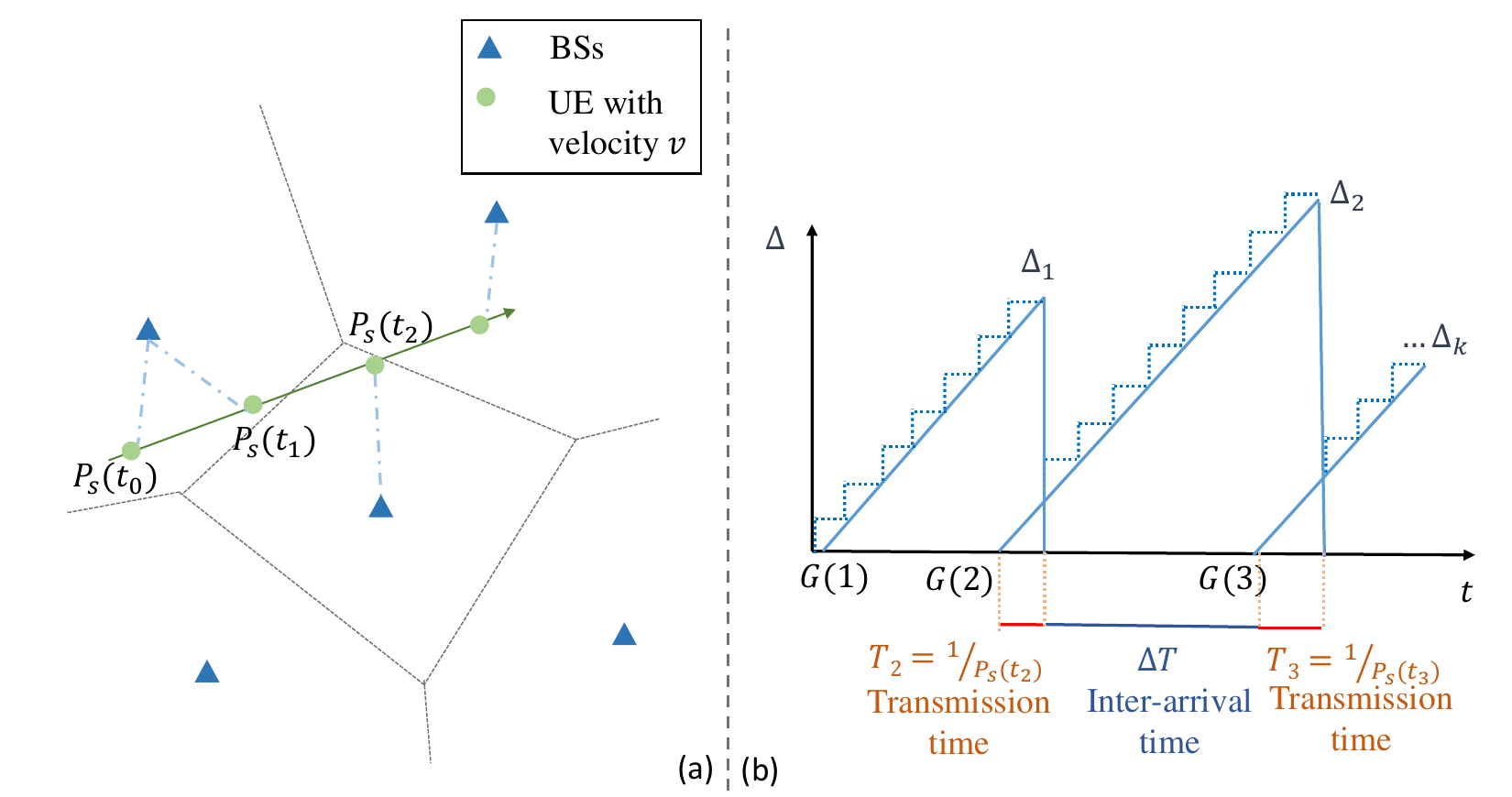}
	\caption{Illustration of the system model.  \textbf{(a)} Illustration of the conditional success probability along the UE trajectory. \textbf{(b)} Illustration of AoI, which is composed of two transmission time, $T_1$ and $T_2$, and one data generation time, $\Delta T$.}
	\label{fig_sysFig}
\end{figure}

We are interested in the spatio-temporal correlation of the conditional success probability of the typical UE, as shown in Fig. \ref{fig_sysFig} (a), which is given in the following definition.
\begin{definition}[Spatio-temporal Correlation Coefficient of $P_s(\theta)$]
The spatio-temporal correlation of the conditional success probability is computed by
\begin{align}
\rho(\theta) = \frac{\mathbb{E}[P_s(\theta,t_0)P_s(\theta,t_1)]-\mathbb{E}^2[P_{s}(\theta)]}{\mathbb{E}[P_{s}^2(\theta)]-\mathbb{E}^2[P_{s}(\theta)]},
\end{align}
where
\begin{align}
	P_s(\theta,t_0)  &= \mathbb{P}({\rm SINR}(t_0) > \theta|\Phi_i(t_0),\Phi),\nonumber\\
	P_s(\theta,t_1)  &= \mathbb{P}({\rm SINR}(t_1) > \theta|\Phi_i(t_1),\Phi),
\end{align}
in which ${\rm SINR}(t_0)$ and ${\rm SINR}(t_1)$ are the SINR at the serving BS of the typical UE at $t_0$ and $t_1$, respectively.
\end{definition}

\subsection{PAoI}
To further capture the impact of the mobility of UE on the performance of the network, we analyze the distribution of the PAoI of the UE, as shown in Fig. \ref{fig_sysFig} (b). We consider the typical device transmitting the updates on the first arrival basis without buffering, new updates can only arrive after the previous transmission success, and the update arrival process is modeled by a Poisson process with arrival rate $\lambda_a$.

Generally, let $\Delta$ be the PAoI and $\bar{\Delta}$ be the mean of the PAoI, which are derived by
\begin{align}
\Delta &= T_{k-1}+Y_{k},\nonumber\\
\bar{\Delta} &= \mathbb{E}[T_{k-1}+Y_{k}\mid \Phi_{i},\Phi],
\end{align}
where $T_k$ is the time of the $k$-th update spent in the system and $Y_{k}$ is the elapsed between the
receptions of the $(k-1)$-th and $k$-th updates. Typically, $T_k$ is the transmission time of an update, and we assume the unit of transmission time is second. 
\begin{align}
	T_k = \frac{1}{P_s(\theta,t_{k})},
\end{align}
and $Y_k$ is computed by the inter-arrival time plus the transmission time
\begin{align}
	Y_k = \frac{1}{P_s(\theta,t_{k})}+T_{\rm tra},
\end{align}
in which $T_{\rm tra}$ is a function of handover delay and update arrival process. While the UE mobility has no influence on the mean PAoI, we analyze the distribution of  PAoI.
\begin{definition}[Distribution of PAoI] The distribution of PAoI is obtained by
\begin{align}
	\label{eq_def_PAoI}
F_{\Delta}(t) &= 	\mathbb{P}(\Delta<t) \nonumber\\
&\approx \mathbb{P}(\mathcal{H})\mathbb{P}(\Delta_{h}<t)+(1-\mathbb{P}(\mathcal{H}))\mathbb{P}(\Delta_{h^{\prime}}<t),
\end{align}
where $\Delta_{h} $ and $\Delta_{h^{\prime}}$ denote the PAoI of UE with and without handover, and the approximation sign comes from the fact that both handover probability and conditional success probability are a function of distance to the serving BS and UE moving direction, however, we take the expectation over the distance and direction separately (i.e., we use the average handover probability as introduced in (\ref{eq_ave_handprob})), and
\begin{align}
\Delta_{h} &= \frac{1}{P_s(\theta,t_{0})}+\frac{1}{P_s(\theta,t_{1})}+t^{\prime}_{\rm tra} ,\nonumber\\
\Delta_{h^{\prime}}
&= 	\frac{1}{P_s(\theta,t_{0})}+\frac{1}{P_s(\theta,t_{1})}+ t_{\rm tra},
\end{align}
where $t_1 = t_0+ t_{\rm tra}$,	the time interval is considered as the average inter-arrival $ t_{\rm tra} = \frac{1}{\lambda_a}$, and $ t^{\prime} =  t_{\rm tra}\times\frac{1}{1-D_{\rm OH}}$, in which $D_{\rm OH}$ denotes the time fraction that the device cannot transfer data due to handover delay \cite{7827020}.
\end{definition}

%
%
%

\section{Performance Analysis}
In this section, we analyze the spatio-temporal correlation coefficient of the conditional success probability for each of ground and aerial UE, and the distribution of PAoI. To do so, we first introduce some important distance distributions, then rewrite SINR meta distribution as functions of the distances to the serving BS and the dominant interferer, finally, we analyze the spatio-temporal correlation of the conditional success probability by computing the correlation of the aforementioned two distances.

\subsection{Ground UE}
As for the uplink communication, BSs are assumed to serve the UE which is located in the Poisson Voronoi (PV) cells, and consequently, the distance to the UE is conditioned on the UE located within the cell. Besides, from the perspective of the typical BS (BS that serves the typical UE), the interference comes from other cells, and the locations of interferers form a non-homogeneous PPP with density $\lambda_i = \lambda\left(1-\exp \left(-\pi \lambda r^2\right)\right)$ \cite{andrews2016primer}, where $r$ is the distance to the typical BS.
We first introduce some important distance distributions of the uplink communication of ground UE, which is given in the following lemma.
\begin{lemma}[Distance Distribution] The probability density function (PDF) and cumulative distribution function (CDF) of $R_0(t_0)$ and the PDF of $R_1(t_0)$ are, respectively, given by 
\begin{align}
	f_{R_0(t_0)}(r) &= 2\pi\lambda r\exp(-\lambda^{\prime}\pi r^2),\nonumber\\
	F_{R_0(t_0)}(r) &= 1-\exp(-\lambda^{\prime}\pi r^2),\nonumber\\
	f_{R_1(t_0)}(r)&=  2 \pi \lambda\left(1-\exp \left(-\pi \lambda r^2\right)\right) r \nonumber\\
	&\times\exp \left(-2 \pi \lambda \int_0^r\left(1-\exp \left(-\pi \lambda z^2\right)\right) z \mathrm{~d} z\right),
\end{align}
where $\lambda^{\prime} = 1.28 \lambda$ is a fitting parameter introduced in \cite{mankar2020distance} due to $R_0(t_0)$ being the distance conditioned on the typical UE located within the PV cell, and  the PDF of $f_{R_1(t_0)}(r)$ follows the properties of non-homogeneous PPP \cite{8833522,9933782}.

Conditioned on the distance to the serving BS at $t_0$ is $R_0(t_0)$, the distance to the serving BS (if no handover happens) at $t_1$ is $R_0(t_1)$, which is shown in Fig. \ref{fig_sys_handover} (a,ii), and its CDF is given as
\begin{align}
	\label{eq_R_0(t_1)|R_0(t_0)}
	F_{R_0(t_1)|R_0(t_0)}(r) &=\frac{\kappa(R_0(t_0),r)}{\pi},
	\end{align}
in which the range of $r$ is $\quad R_0(t_0)-vt<r<R_0(t_0)+vt$, and
\begin{align}
	\label{eq_kappa}
	\kappa(R_0(t_0),r) &= \arccos\bigg(\frac{(vt)^2+R_{0}^{2}(t_0)-r^2}{2vtR_{0}(t_0)}\bigg).
\end{align}
\end{lemma}
\begin{IEEEproof}
	(\ref{eq_R_0(t_1)|R_0(t_0)}) is obtained by
\begin{align}
&	F_{R_0(t_1)|R_0(t_0)}(r)= \mathbb{P}(R_0(t_1)<r)\nonumber\\
	&= \mathbb{P}\bigg(\sqrt{(vt)^{2}+R_{0}^{2}(t_0)-2vtR_{0}(t_0)\cos(\kappa_0)}<r\bigg),\nonumber
\end{align}	
the relations of $R_0(t_1)$ and $R_0(t_0)$ are shown in Fig. \ref{fig_sys_handover} (a,ii).
\end{IEEEproof}

The spatio-temporal correlation of the conditional success probability comes from the fact that the locations of the interferers and typical UE are correlated. While beta approximation is commonly used in computing the distribution of conditional success probability, it is difficult to use in analyzing the spatio-temporal correlation, therefore, we use the dominant interferer-based approximation proposed in \cite{10066317}. The main idea of this approximation is to consider the dominant interferer, which is the nearest LoS or NLoS UE that provides the strongest average signal, exactly, while treating the other interferers in an average sense.
Firstly, we approximate the interference of the uplink communication.

\begin{lemma}[Approximation of the Interference.]
The uplink communication interference of ground UE is approximated by
\begin{align}
	I_1(R_1) &= H_{x_1} p_t R_{1}^{-\alpha}+I^{\prime}(R_1),
\end{align}
where \footnote{Note that throughout this paper, we might in some particular situations drop the argument of $R_0$ or $R_1$ for simplicity of notation.}
\begin{align}
	I^{\prime}(R_1) = 2\pi\lambda p_t\frac{x^{2-\alpha}}{\alpha-2}-p_t (\lambda\pi)^{\alpha/2}\Gamma\bigg(1-\frac{\alpha}{2},\lambda\pi x^2\bigg),
\end{align}
where $\Gamma(a,z) = \int_{z}^{\infty}t^{a-1}\exp(-t){\rm d}t$.
\end{lemma}
\begin{IEEEproof}
Similar to \cite{10066317,kishk2017joint}, we considering the interferernce is approximated by consider the dominant term, (the nearest interferer in ground UE scenario), exactly, while the mean of the remaining interferers,
	\begin{align}
		I &= \sum_{x\in\Phi_{i}} H_x p_t D_{i,x}^{-\alpha} = H_{x_1} p_t R_{1}^{-\alpha}+\sum_{x\in\Phi_{i}\setminus x_1} H_x p_t D_{i,x}^{-\alpha}\nonumber\\
		&\approx H_{x_1} p_t R_{1}^{-\alpha}+I^{\prime}(R_1),\nonumber\\
		I^{\prime}(R_1) &=\mathbb{E}\bigg[\sum_{x\in\Phi_{i}\setminus x_1} H_x p_t D_{i,x}^{-\alpha}\bigg]= 2\pi p_t \int_{R_1}^{\infty} \lambda_i(z) z^{-\alpha+1}{\rm d}z \nonumber\\
		&= 2\pi p_t \int_{R_1}^{\infty} \lambda(1-\exp(-\pi \lambda z^2)) z^{-\alpha+1}{\rm d}z 
		\nonumber\\
		&= 2\pi\lambda p_t\frac{x^{2-\alpha}}{\alpha-2}-p_t (\lambda\pi)^{\alpha/2}\Gamma\bigg(1-\frac{\alpha}{2},\lambda\pi x^2\bigg),
	\end{align}
where $\Phi_{i}$ denotes the locations of the interferers: $\Phi_{i}\in\Phi_{u}$.
\end{IEEEproof}

In what follows, we derive the joint distribution of the conditional success probability for ground UE. By using the results from \cite{10066317} directly, the approximated conditional success probability is given by
\begin{align}
P_{s}(\theta,t_0) \approx  \exp\biggl\{-\frac{\theta(I_1(R_{1}(t_0))+\sigma^2)}{p_t R_{0}(t_0)}\biggr\}\frac{1}{1+\theta \left(\frac{R_{0}(t_0)}{R_{1}(t_0)}\right)^{\alpha}},
\end{align}
 and the joint distribution of the conditional success probability 
$\mathbb{P}(P_s(\theta,t_0)>\gamma_0,P_s(\theta,t_1)>\gamma_1)$ can be represented by using the conditional probability as follows:
\begin{align}
	&\mathbb{P}(P_s(\theta,t_0)>\gamma_0,P_s(\theta,t_1)>\gamma_1) \nonumber\\
	&= \mathbb{P}(P_s(\theta,t_1)>\gamma_1 \mid P_s(\theta,t_0)>\gamma_0 )\mathbb{P}(P_s(\theta,t_0)>\gamma_0),
\end{align}
and solved by considering the spatial correlation between $t_0$ and $t_1$, that is the correlations between $R_0(t_0)$ and $R_0(t_1)$, and $R_1(t_0)$ and $R_1(t_1)$, as given in the following theorem.

\begin{theorem}[Joint Distribution of the Conditional Success Probability]\label{theo_JointPs_GU}
In the case of no handover, the joint distribution of the conditional success probability is computed by
\begin{align}
	&F_{P_{s_0},P_{s_1},h^{\prime}}(\gamma_0,\gamma_1) 
	 =\int_{0}^{\infty}\int_{K(r_1,\gamma_0,\theta)}^{\infty}\int_{0}^{\pi}\nonumber\\
	 &\quad\frac{1-\kappa(r_0,K(g^{-1}(r_1),\gamma_1,\theta))}{\pi^2}{\rm d}\kappa_1f_{R_0(t_0)}^{\prime}(r_0){\rm d}r_0 f_{R_1(t_0)}(r_1){\rm d}r_1\nonumber\\
	 &\times\int_{0}^{\infty} 1-F_{R_0(t_0)}(K(r_1,\gamma_0,\theta))  f_{R_1(t_0)}(r_1){\rm d}r_1,
\end{align}
where $K(r,\gamma,\theta) = \left(-\frac{r^{\alpha}}{\theta }+\frac{1}{s(r)}W\left(0,\frac{s(r)\exp(s(r)\theta^{-1}r^\alpha)}{\gamma\theta r^{-\alpha}}\right)\right)^{\frac{1}{\alpha}}$, $s(r) = \frac{\theta}{p_t}(I_1(r)+\sigma^2)$, $\kappa(r_0,r_1)$ is given in (\ref{eq_kappa}),  $g(\cdot)$ is the function defined in (\ref{eq_R0R1}), $g^{-1}(\cdot)$ is the inverse function, and
\begin{align}
	f_{R_{0}(t_0)}^{\prime}(r)&= \frac{f_{R_{0}(t_0)}(r)}{\exp(-\pi\lambda K^{2}(R_{1}(t_0),\gamma_0,\theta) )}.
\end{align}

In the case of handover, the joint distribution of the conditional success probability is computed by
\begin{align}
	&F_{P_{s_0},P_{s_1},h}(\gamma_0,\gamma_1) 
	=\int_{0}^{\infty}\int_{0}^{\infty}\int_{0}^{\pi}\int_{K(z_0,\gamma_0,\theta)}^{\infty}\nonumber\\
	&\quad 1- F_{R_{0}^{\prime}|\kappa_0,R_0}(K(z_1,\gamma_1,\theta),\kappa,r_0)\frac{1}{\pi}{\rm d}\kappa f_{R_1(t_0)}(z_1){\rm d}z_1 \nonumber\\
	&\times f_{R_0(t_0)}(r_0){\rm d}r_0 f_{R_1(t_0)}(z_0){\rm d}z_0\nonumber\\
	&\times\int_{0}^{\infty} 1-F_{R_0(t_0)}(K(r_1,\gamma_0,\theta))  f_{R_1(t_0)}(r_1){\rm d}r_1.
\end{align}

\end{theorem}
\begin{IEEEproof}
	See Appendix \ref{app_JointPs_GU}.
	\end{IEEEproof}

\subsection{Aerial UE}
In this subsection, we analyze the correlation performance for aerial UEs, such as UAVs. Similar to ground UEs, BSs are assumed to serve the aerial UEs in their PV cells, and the locations of interferers are modeled by a non-homogeneous PPP $\Phi_{i}$  with density $\lambda_i = \lambda\left(1-\exp \left(-\pi \lambda r^2\right)\right)$, where $r$ is the distance to the typical BS. 

Different from ground UEs, aerial UEs have higher chances of establishing LoS linkswith BSs due to their high altitude. Therefore, the uplink communication interference is divided into, LoS interferers and NLoS interferers:
\begin{align}
	I &= \sum_{u_i\in\Phi_{i}^{l}} p_t \eta_{ l} G_l R_{u_i}^{-\alpha_{l}}+\sum_{u_i\in\Phi_{i}^{n}} p_t\eta_{n} G_n R_{u_i}^{-\alpha_{n}},\label{eq_I}
\end{align}
in which $\Phi_{i}^{l}$ and $\Phi_{i}^{n}$ are subsets of $\Phi_{i}$ denoting the locations of interfering users which establish LoS/NLoS links with the typical BS.

We first introduce some important distance distributions which are needed for the correlation analysis. While the distribution of the distance to the serving BS is the same as ground UEs, that of the distance to the dominant interferer is different due to the non-homogenity of the locations of LoS and NLoS interferers.
\begin{lemma}[Distance Distribution] Let $R_{l}$ and $R_{n}$ be the distances from the typical BS to the nearest LoS and NLoS interferer, the PDF and CDF of $R_{l}$ and $R_{n}$ are given by, respectively,
	\begin{align}
		\label{eq_au_FRnFRl}
		f_{R_{l}}(r) &= 2\pi r\lambda\left(1-\exp \left(-\pi \lambda r^2\right)\right) P_l(r)\nonumber\\
		&\times\exp\bigg(-\int_{0}^{r}2\pi z\lambda\left(1-\exp \left(-\pi \lambda z^2\right)\right) P_l(z){\rm d}z\bigg),\nonumber\\
		F_{R_{l}}(r) &= 1-\exp\bigg(-\int_{0}^{r}2\pi z\lambda\nonumber\\
		&\qquad\times\left(1-\exp \left(-\pi \lambda z^2\right)\right) P_l(r){\rm d}z\bigg),\nonumber\\
		f_{R_{n}}(r) &= 2\pi r\lambda\left(1-\exp \left(-\pi \lambda r^2\right)\right) P_n(r)\nonumber\\
		&\times\exp\bigg(-\int_{0}^{r}2\pi z\lambda\left(1-\exp \left(-\pi \lambda z^2\right)\right) P_n(z){\rm d}z\bigg),\nonumber\\
		F_{R_{n}}(r) &= 1-\exp\bigg(-\int_{0}^{r}2\pi z\lambda\nonumber\\
		&\qquad\times\left(1-\exp \left(-\pi \lambda z^2\right)\right) P_n(z){\rm d}z\bigg).
	\end{align}
\end{lemma}

Similar to the ground UE case, we consider the dominant interferer exactly, while the remaining interferers in average. The dominant interferer now has two scenarios, LoS dominant interferer and NLoS dominant interferer. Therefore, there are two integrals, one for LoS interference and one for NLoS interference, when we compute the mean of the remaining interference.  The  approximated interference is given in the following lemma.
\begin{lemma}[Approximated Interference for Aerial UE] For the aerial UE, the interference of the uplink communication can be approximated by
\begin{align}
	I_l(R_{l}) &\approx p_t \eta_l G_l (R_{l}^2+h^2)^{-\frac{\alpha_l}{2}}+ s_l(R_{l}),\nonumber\\
	s_l(r) &= 2\pi p_t\eta_l\int_{R_{l}}^{\infty}  P_l(r)\lambda_i (r^2+h^2)^{-\frac{\alpha_l}{2}}r{\rm d}r \nonumber\\
	&+2\pi p_t\eta_n\int_{d_{ln(R_{l})}}^{\infty}  P_n(r)\lambda_i(r^2+h^2)^{-\frac{\alpha_n}{2}}r{\rm d}r,\\
	I_n(R_{n}) &\approx p_t\eta_n  G_n (R_{n}^2+h^2)^{-\frac{\alpha_n}{2}}+ s_n(R_{n}),\nonumber\\
	s_n(r) &= 2\pi p_t\eta_l\int_{d_{nl(R_{n})}}^{\infty}  P_l(r)\lambda_i(r^2+h^2)^{-\frac{\alpha_l}{2}}r{\rm d}r \nonumber\\
	&+2\pi p_t\eta_n\int_{R_{n}}^{\infty}  P_n(r)\lambda_i(r^2+h^2)^{-\frac{\alpha_n}{2}}r{\rm d}r,
\end{align}
where $d_{ln}(r)$ denotes the distance to the nearest NLoS interferer given  the dominant interferer is a LoS aerial UE and it is located at $r$ away, similar to $d_{nl}(r)$,
\begin{align}
	d_{ln}(r) &= \sqrt{\bigg(\max\bigg(h,\bigg(\frac{\eta_n}{\eta_l}\bigg)^{1/\alpha_{n}}r^{\alpha_{l}/\alpha_n}\bigg)\bigg)^2-h^2},\nonumber\\ 
	d_{nl}(r) &= \sqrt{\bigg(\bigg(\frac{\eta_l}{\eta_n}\bigg)^{1/\alpha_{l}}r^{\alpha_{n}/\alpha_l}\bigg)^2-h^2}.
\end{align}
\end{lemma}

Now we proceed to analyze the spatio-temporal correlation of aerial UE. We  first derive the approximated conditional success probability in the following lemma.
\begin{lemma}[Approximated Conditional Success Probability of Aerial Users] 
	\label{lemma_condPs_au}
	By using the dominant interferer-based approximation, the conditional success probability is computed as
\begin{align}
	\label{eq_au_Ps}
P_s(\theta) \approx P_{s,ll}(\theta)+P_{s,ln}(\theta)+P_{s,nl}(\theta)+P_{s,nn}(\theta),
\end{align}
where,
\begin{align}
P_{s,ll}(\theta) &= \mathbbm{1}({\rm LoS}) \mathbbm{1}(R_l)	\kappa_{ll}(R_0,R_{l}),\nonumber\\
P_{s,ln}(\theta) &= \mathbbm{1}({\rm LoS}) \mathbbm{1}(R_n)	\kappa_{ln}(R_0,R_{n}),\nonumber\\
P_{s,nl}(\theta) &= \mathbbm{1}({\rm NLoS}) \mathbbm{1}(R_l)	\kappa_{nl}(R_0,R_{l}),\nonumber\\
P_{s,nn}(\theta) &= \mathbbm{1}({\rm NLoS}) \mathbbm{1}(R_n)	\kappa_{nn}(R_0,R_{n}),
\end{align}
in which,
\begin{small}
\begin{align}
	\kappa_{ll}(r,r_{l}) &= \sum_{k = 1}^{m_l}\binom{m_l}{k}(-1)^{k+1}\exp\bigg(-\frac{k\beta_l m_l \theta(I_{l}(r_{l})+\sigma^2)}{p_t \eta_l (r^2+h^2)^{-\frac{\alpha_{l}}{2}}}\bigg)\nonumber\\
	&\times\left(1+k\beta_l\theta (r_{l}^2+h^2)^{\frac{-\alpha_{l}}{2}}(r^2+h^2)^{\frac{\alpha_{l}}{2}}\right)^{-m_l},\nonumber\\
	\kappa_{ln}(r,r_{n}) &= \sum_{k = 1}^{m_l}\binom{m_l}{k}(-1)^{k+1}\nonumber\\
	&\times\exp\bigg(-\frac{k\beta_l m_l \theta(I_{n}(r_n)+\sigma^2)}{p_t \eta_l (r^2+h^2)^{-\frac{\alpha_{l}}{2}}}\bigg)\nonumber\\
	&\times\left(1+k\beta_l\theta m_l (r_{n}^{2}+h^2)^{-\frac{\alpha_{n}}{2}}(r^2+h^2)^{\frac{\alpha_{l}}{2}}\eta_n\right)^{-m_n},\nonumber\\
	\kappa_{nl}(r,r_{l}) &= \exp\bigg(-\frac{\theta(I_{l}(r_l)+\sigma^2)}{p_t \eta_n (r^2+h^2)^{\frac{\alpha_{n}}{2}}}\bigg)\nonumber\\
	&\times m_{l}^{m_{l}}\left(m_{l}+\theta \eta_{l}(r_{l}^2+h^2)^{-\frac{\alpha_{l}}{2}}\eta_{n}^{-1}R_{u}^{\alpha_{n}}\right)^{-m_l},\nonumber\\
	\kappa_{nn}(r,r_{n}) &= \exp\bigg(-\frac{ \theta(I_{n}(r_n)+\sigma^2)}{p_t \eta_n (r^2+h^2)^{-\frac{\alpha_{n}}{2}}}\bigg)\nonumber\\
	&\times\left(1+\theta (r_{n}^{2}+h^2)^{-\frac{\alpha_{n}}{2}}(r^2+h^2)^{\frac{\alpha_{n}}{2}}\right).
\end{align}
\end{small}
\end{lemma}
\begin{IEEEproof}
	See Appendix \ref{app_condPs_au}.
\end{IEEEproof}

Now the communication links between the typical BS and each of  the typical user and the dominant interfering user are divided into four scenarios: establish LoS/NLoS link with the typical user while the dominant interferer is a LoS/NLoS user. Let $\mathcal{A}_{ll}(R_0,R_{l})$, $\mathcal{A}_{ln}(R_0,R_{n})$, $\mathcal{A}_{nl}(R_0,R_{l})$, and $\mathcal{A}_{nn}(R_0,R_{n})$ denote the association events mentioned above, the probabilities of these events are computed in the following lemma.
\begin{lemma}[Association Probability] The association probabilities of $\mathcal{A}_{ll}(R_0,R_{l})$, $\mathcal{A}_{ln}(R_0,R_{n})$, $\mathcal{A}_{nl}(R_0,R_{l})$, and $\mathcal{A}_{nn}(R_0,R_{n})$ are
	\begin{align}
		\mathcal{A}_{ll}(R_0,R_{l}) &= P_l(R_0)\bar{F}_{R_n}(d_{ln}(R_{l})),\nonumber\\	
		\mathcal{A}_{ln}(R_0,R_{n}) &= P_l(R_0)\bar{F}_{R_l}(d_{nl}(R_{n})),\nonumber\\	
		\mathcal{A}_{nl}(R_0,R_{l}) &= P_n(R_0)\bar{F}_{R_n}(d_{ln}(R_{l})),\nonumber\\	
		\mathcal{A}_{nn}(R_0,R_{n}) &= P_n(R_0)\bar{F}_{R_l}(d_{nl}(R_{n})),
	\end{align}	
	where $\bar{F}_{\{\cdot\}}(r)$ is CCDF obtained by $1-F_{\{\cdot\}}(r)$ mentioned in (\ref{eq_au_FRnFRl}).
\end{lemma}
\begin{IEEEproof}
	While the probability that the UAV establishes a LoS link with the typical user located at a horizontal distance $R_0$ away is $P_l(R_0)$, the probability of the dominant interferer is a LoS user at horizontal distance $R_{l}$ is the probability that the nearest NLoS user should be at least  located at $d_{nl}(\sqrt{R_{l}^2+h^2})$ away,
	\begin{align}
		&\mathcal{A}_{ll}(R_0,R_{l}) \nonumber\\
		&= P_l(R_0)\times\mathbb{P}(\eta_{ l}(R_{l}^2+h^2)^{-\frac{\alpha_l}{2}}>\eta_{ n}(R_{n}^2+h^2)^{-\frac{\alpha_n}{2}}).
	\end{align}
\end{IEEEproof}

\begin{theorem}[Approximated SINR Meta Distribution of Aerial Users] By using the dominant interferer-based approximation, the SINR meta distribution of aerial users is given by
	\begin{align}
		&\bar{F}_{P_s}(\theta,\gamma)  \approx \iint_{\mathbbm{1}(\kappa_{ll}(r,r_{l})>\gamma)}	\mathcal{A}_{ll}(r,r_{l})f_{R_0}(r)f_{R_{l}}(r_l){\rm d}r_l{\rm d}r \nonumber\\
		&\quad+\iint_{\mathbbm{1}(\kappa_{ln}(r,r_{n})>\gamma)}\mathcal{A}_{ln}(r,r_{n})f_{R_0}(r)f_{R_n}(r_n){\rm d}r_n{\rm d}r\nonumber\\
		&\quad+\iint_{\mathbbm{1}(\kappa_{nl}(r,r_{l})>\gamma)}\mathcal{A}_{nl}(r,r_{n})f_{R_0}(r)f_{R_l}(r_l){\rm d}r_l{\rm d}r  \nonumber\\
		&+\iint_{\mathbbm{1}(\kappa_{nn}(r,r_{n})>\gamma)}\mathcal{A}_{nn}(r,r_{n})f_{R_0}(r)f_{R_{n}}(r_n){\rm d}r_n{\rm d}r.
	\end{align}
\end{theorem}
\begin{IEEEproof}
	Since we use indicator functions in (\ref{eq_au_Ps}), the SINR meta distribution is derived by:
	\begin{align}
		&\quad\bar{F}_{P_s}(\theta,\gamma) = \mathbb{P}(P_s(\theta)>\gamma) \nonumber\\
		&= \mathbb{P}([P_{s,ll}(\theta)+P_{s,ln}(\theta)+P_{s,nl}(\theta)+P_{s,nn}(\theta)]>\gamma)\nonumber\\
		&\approx\mathbb{E}_{\{R_{0},R_{l},R_{n}\}}[\mathcal{A}_{ll}(R_0,R_{l})\mathbb{P}( \kappa_{ll}(R_0,R_{l})>\gamma)\nonumber\\
		&\quad+\mathcal{A}_{ln}(R_0,R_{n}) \mathbb{P}(	\kappa_{ln}(R_0,R_{n})>\gamma)\nonumber\\
		&\quad+\mathcal{A}_{nl}(R_0,R_{l})\mathbb{P}(\kappa_{nl}(R_0,R_{l})>\gamma)\nonumber\\
		&\quad+\mathcal{A}_{nn}(R_0,R_{n})\mathbb{P}(	\kappa_{nn}(R_0,R_{n}) >\gamma)],
	\end{align}
	proof completes by integrating over $R_0,R_{l}$, and $R_{n}$.
\end{IEEEproof}
\begin{remark}
	The reason for using  indicator functions is that: compared with the Rayleigh fading model, Nakagami-m fading is more complex since the power of LoS channel fading follows a Gamma distribution with shape and scale parameter $(m_l,1/m_l)$. Consequently, step (a) in \cite[Eq. (21)]{10066317} cannot be solved by using the Lambert W function.
	
	Compared with the beta approximation, which is computed in \cite{10218328,10024838}, the proposed method with indicator function does not require computing the moments of the conditional success probability, given in \cite[Theorem 2]{10218328}, as well as the Laplace transform of the interference, given in \cite[Eq. (44)-Eq. (47)]{10218328}, which is the most time-consuming part of the analysis, and thus, the proposed method highly reduces the computation complexity. 
\end{remark}

Consequently, the joint distribution of the conditional success probability of the aerial users, which is similar to the ground user scenario: rewrite the joint distribution by considering the correlations between $R_0(t_0)$ and $R_0(t_1)$, and $R_1(t_0)$ and $R_1(t_1)$, is derived in the following theorem.
\begin{theorem}[Joint Distribution of the Conditional Success Probability of Aerial Users]
	\label{theorem_JointPs_au}
In the case of no handover, the joint distribution of the conditional success probability is computed by
\begin{align}
	&F_{P_{s_0},P_{s_1},h^{\prime}}(\gamma_0,\gamma_1) =\sum_{i_0,j_0,i_1,j_1\in\{l,n\}}\int_{0}^{\pi}\iiint_{\mathbbm{1}(c_{h}^{\prime} )}\frac{1}{\pi}\nonumber\\
	&\qquad\times \mathcal{A}_{i_0j_0}(r_0,z_0)\mathcal{A}_{i_1j_1}(g(r_0,\kappa),z_1)\nonumber\\
	&\qquad\times f_{R_0(t_0)}(r_0){\rm d}r_0 f_{R_{j_0}}(z_0){\rm d}z_0 f_{R_{j_1}}(z_1){\rm d}z_1{\rm d}\kappa,
\end{align}
where $c_{h}^{\prime} = \kappa_{i_0j_0}(r_0,z_0)>\gamma_0, \kappa_{i_1j_1}(g(r_0,\kappa_0),z_1)>\gamma_1$.

In the case of handover, the joint distribution of the condtional success probability is computed by
\begin{align}
	&F_{P_{s_0},P_{s_1},h}(\gamma_0,\gamma_1) \approx \sum_{i_0,j_0,i_1,j_1\in\{l,n\}}\iiiint_{\mathbbm{1}(c_{h} )}\mathcal{A}_{i_0j_0}(r_0,z_0)\nonumber\\
	&\times \mathcal{A}_{i_1j_1}(r_1,z_1) f_{R_1|\kappa_0,R_0}(r_1,\pi,r_0)f_{R_0(t_0)}(r_1){\rm d}r_1\nonumber\\
	&\times  f_{R_0(t_0)}(r_0){\rm d}r_0 f_{R_{j_0}}(z_0){\rm d}z_0 f_{R_{j_1}}(z_1){\rm d}z_1,
\end{align}
where $c_{h} = \kappa_{i_0j_0}(r_0,z_0)>\gamma_0, \kappa_{i_1j_1}(r_1,z_1)>\gamma_1$, and the approximation sign follows from we set $\kappa_0 = \pi$ to reduce the number of integrations.
\end{theorem}
\begin{IEEEproof}
	See Appendix \ref{app_JointPs_au}.
\end{IEEEproof}
\begin{remark}
Compared to the scenario involving ground UEs, the joint distribution of conditional success probability becomes significantly more complex in the case of aerial UEs. This complexity arises from two key factors:
(i) the dominant interferer in the aerial scenario may be the nearest Line of Sight (LoS) or Non-Line of Sight (NLoS) aerial UE, and (ii) the locations of LoS and NLoS interferers change at two different time instants. In other words, interferer $x_i$ might establish a LoS link at time $t_0$ and an NLoS link with the tagged BS at time $t_1$.
Considering these factors, we can anticipate that the correlation coefficient of the conditional success probability for aerial UEs will be lower than that of ground UEs. This observation is also supported by the numerical results presented in the subsequent sections.
\end{remark}

\subsection{Correlation}

In this subsection, we compute the spatio-temporal correlation of ground and aerial UE, respectively. 
We first derive the PDF of the joint distribution of the conditional success probability. While the PDF should be derived by taking the derivative of the CDF,
\begin{align}
f_{P_{s0},P_{s1},h}(\gamma_0,\gamma_1) = \frac{\partial^2  F_{P_{s_0},P_{s_1},h}(\gamma_0,\gamma_1)}{\partial \gamma_0 \partial\gamma_1},
\end{align}
it is difficult to obtain the exact expression due to the indicator functions in the analysis for aerial UE, and the complexity of the joint CDF of the conditional success probability of both ground and aerial UE. Therefore, we obtain the PDF numerically, which is given in the following lemma.

\begin{lemma}[Joint PDF of the Conditional Success Probability]
$f_{P_{s0},P_{s1},\{h,h^{\prime}\}}(\gamma_0,\gamma_1)$ is numerically discretized by
\begin{align}
	&f_{P_{s0},P_{s1},\{h,h^{\prime}\}}(\gamma_0,\gamma_1)\approx \frac{1}{\delta^2}(F_{P_{s0},P_{s1},\{h,h^{\prime}\}}(\gamma_0,\gamma_1) \nonumber\\
	& -F_{P_{s0},P_{s1},\{h,h^{\prime}\}}(\gamma_0-\delta,\gamma_1)-F_{P_{s0},P_{s1},\{h,h^{\prime}\}}(\gamma_0,\gamma_1-\delta)\nonumber\\
	&+F_{P_{s0},P_{s1},\{h,h^{\prime}\}}(\gamma_0-\delta,\gamma_1-\delta)),
\end{align}
where $\delta$ is set at $0.01$.
\end{lemma}

In what follows, we compute the spatio-temporal correlation coefficient of the conditional success probability, which requires computing the first and second moments of the conditional success probability. These two moments are obtained by taking the expectation over the random variable  (which is $R_1$ in ground UE scenario, and $R_l$ and $R_n$ in aerial UE scenario): $M_1(\theta) = \mathbb{E}[P_{s}(\theta)]$ and $M_2(\theta) = \mathbb{E}[P_{s}^{2}(\theta)]$.

\begin{theorem}[Spatio-temporal Correlation] The spatio-temporal correlation coefficient of ground and aerial UE is obtained by
\begin{align}
	\rho(\theta) = \frac{{\rm Cov}}{M_{2}(\theta)-M_{1}^{2}(\theta)},
\end{align}
where,
\begin{align}
&{\rm Cov} \approx 	(1-\mathbb{P}(\mathcal{H}))\iint \gamma_0\gamma_1 f_{P_{s0},P_{s1},h^{\prime}}(\gamma_0,\gamma_1) {\rm d}\gamma_0{
		\rm d}\gamma_1\nonumber\\
	&+	\mathbb{P}(\mathcal{H})\iint \gamma_0\gamma_1 f_{P_{s0},P_{s1},h}(\gamma_0,\gamma_1) {\rm d}\gamma_0{
		\rm d}\gamma_1-M_{1}^{2}(\theta),
\end{align}
where $M_{1}(\theta)$ and $M_{2}(\theta)$ are the first and the second moments of the conditional success probability. As for the ground UE, $M_{1}(\theta)$ and $M_{2}(\theta)$ are derived by
\begin{align}
M_{1}^{2}(\theta) &\approx \left(\int_{0}^{\infty}\int_{0}^{r_1}\frac{2r_0}{r_{1}^2}f_{r_1}(r_1)P_{s}(\theta,t_0){\rm d}r_0{\rm d}r_1\right)^2,\nonumber\\
M_{2}(\theta) &\approx \int_{0}^{\infty}\int_{0}^{r_1}\frac{2r_0}{r_{1}^2}f_{r_1}(r_1)P_{s}^{2}(\theta,t_0){\rm d}r_0{\rm d}r_1,
\end{align}
where $f_{r_1}(r) = 2(\pi\lambda)^2 r^3 \exp(-\pi\lambda r^2)$. As for the aerial UE, $M_{1}(\theta)$ and $M_{2}(\theta)$ are derived by
	\begin{align}
	&M_{1}^{2}(\theta)  \approx \bigg[\int_{0}^{\infty}\int_{0}^{\infty}	\mathcal{A}_{ll}(r,r_{l})\kappa_{ll}(r,r_{l}) f_{R_0}(r)f_{R_{l}}(r_l){\rm d}r_l{\rm d}r \nonumber\\
	&\quad+\int_{0}^{\infty}\int_{0}^{\infty}\mathcal{A}_{ln}(r,r_{n})\kappa_{ln}(r,r_{n})f_{R_0}(r)f_{R_n}(r_n){\rm d}r_n{\rm d}r\nonumber\\
	&\quad+\int_{0}^{\infty}\int_{0}^{\infty}\mathcal{A}_{nl}(r,r_{n})\kappa_{nl}(r,r_{l})f_{R_0}(r)f_{R_l}(r_l){\rm d}r_l{\rm d}r  \nonumber\\
	&+\int_{0}^{\infty}\int_{0}^{\infty}\mathcal{A}_{nn}(r,r_{n})\kappa_{nn}(r,r_{n})f_{R_0}(r)f_{R_{n}}(r_n){\rm d}r_n{\rm d}r\bigg]^2,\nonumber\\
	&M_{2}(\theta)  \approx \int_{0}^{\infty}\int_{0}^{\infty}	\mathcal{A}_{ll}(r,r_{l})\kappa_{ll}^2(r,r_{l}) f_{R_0}(r)f_{R_{l}}(r_l){\rm d}r_l{\rm d}r \nonumber\\
	&\quad+\int_{0}^{\infty}\int_{0}^{\infty}\mathcal{A}_{ln}(r,r_{n})\kappa_{ln}^2(r,r_{n})f_{R_0}(r)f_{R_n}(r_n){\rm d}r_n{\rm d}r\nonumber\\
	&\quad+\int_{0}^{\infty}\int_{0}^{\infty}\mathcal{A}_{nl}(r,r_{n})\kappa_{nl}^2(r,r_{l})f_{R_0}(r)f_{R_l}(r_l){\rm d}r_l{\rm d}r  \nonumber\\
	&+\int_{0}^{\infty}\int_{0}^{\infty}\mathcal{A}_{nn}(r,r_{n})\kappa_{nn}^2(r,r_{n})f_{R_0}(r)f_{R_{n}}(r_n){\rm d}r_n{\rm d}r
\end{align}
in which the approximation signs come from the fact that we use the dominant interferer-based approximation to compute the conditional success probability.
\end{theorem}

\begin{remark}
	When the velocity of UE approaches infinity, the correlation coefficient drops to zero:
	\begin{align}
	&	{\rm Cov}_{\infty} \stackrel{(a)}{=} \mathbb{P}(\mathcal{H})\iint \gamma_0\gamma_1 f_{P_{s0},P_{s1},h}(\gamma_0,\gamma_1) {\rm d}\gamma_0{\rm d}\gamma_1-M_{1}^{2}(\theta)\nonumber\\
		&\stackrel{(b)}{=} \int_{0}^{1} \gamma_0 f_{P_{s0}}(\gamma_0) {\rm d}\gamma_0  \int_{0}^{1} \gamma_1 f_{P_{s1}}(\gamma_1) {\rm d}\gamma_1-M_{1}^{2}(\theta) = 0,\nonumber
	\end{align}
in which step (a) follows from that the handover must occur, and step (b) is because $R_{d}$ approaches $\infty$, and the distribution of $R_{0}^{\prime}(t_1)$ becomes uncorrelated with $R_{0}(t_0)$, as shown in Fig. \ref{fig_sys_handover} (b), then (\ref{eq_57}) can be computed separately as $F_{R^{\prime}_{0}|{\kappa_{0},R_0}}(r,\kappa_{0},R_0) = 1-\exp(-\pi r\lambda)$ now.
	\end{remark}

\subsection{Age of Information}
In this subsection, we encompass UE mobility effect and compute the PAoI of ground and aerial UE. Let $R_{\rm h}(v)$ denote  the average number of handovers per unit time, which is the average number of intersections between UE trajectory and cell boundaries per unit time, given the velocity of UE is $v$ \cite{7827020}.   Besides, we consider a handover delay for each handoff during which no data can be transmitted. 
\begin{align}
	{R}_{\rm h}(v) &= \frac{v\lambda^{\frac{1}{2}}}{\pi}\int_{0}^{\pi}\sqrt{2-2\cos(\theta)}{\rm d}\theta, \nonumber\\
	{\rm D_{HO}} &= \min(1,d_m {R}_{\rm h}(v)),
\end{align}
where $d_m$ is the handover delay, and ${\rm D_{HO}}$ is the fraction of time that the device cannot transfer massage, and the velocity of UE is assumed to guarantee $D_{\rm HO}<1$.

\begin{theorem}
The distribution of PAoI, which is defined in (\ref{eq_def_PAoI}), is obtained by
\begin{align}
&	F_{\Delta}(t) \approx \mathbb{P}(\mathcal{H})\iint \kappa_{aoi,h}(\gamma_0,\gamma_1)f_{P_{s0},P_{s1},h}(\gamma_0,\gamma_1){\rm d}\gamma_0{\rm d}\gamma_1\nonumber\\
	& +(1-\mathbb{P}(\mathcal{H}))\iint \kappa_{aoi,h^{\prime}}(\gamma_0,\gamma_1)f_{P_{s0},P_{s1},h^{\prime}}(\gamma_0,\gamma_1){\rm d}\gamma_0{\rm d}\gamma_1,
\end{align}
where $t_{\rm tra} = \frac{1}{\lambda_a}$,
\begin{align}
\kappa_{aoi,h}(\gamma_0,\gamma_1) &= \mathbbm{1}\bigg( \bigg[\frac{1}{\gamma_0}+\frac{1}{\gamma_1}+ \frac{t_{\rm tra}}{1-D_{\rm OH}}\bigg]<t\bigg),\nonumber\\
\kappa_{aoi,h^{\prime}}(\gamma_0,\gamma_1) &=  \mathbbm{1}\bigg( \bigg[\frac{1}{\gamma_0}+\frac{1}{\gamma_1}+ t_{\rm tra}  \bigg]<t\bigg).
\end{align}
\end{theorem}
\begin{IEEEproof}
The distribution of the PAoI is obtained by
\begin{align}
&\mathbb{P}(\Delta<t) \approx \mathbb{P}(\mathcal{H})\mathbb{P}(\Delta_{h}<t)+(1-\mathbb{P}(\mathcal{H}))\mathbb{P}(\Delta_{h^{\prime}}<t)\nonumber\\
&= \mathbb{P}(\mathcal{H})\mathbb{P}\bigg(\bigg[\frac{1}{P_s(\theta,t_{0})}+\frac{1}{P_s(\theta,t_{0}+t_{\rm tra})}+ \frac{t_{\rm tra}}{1-D_{\rm OH}}\bigg]<t\bigg)\nonumber\\
&+(1-\mathbb{P}(\mathcal{H}))\mathbb{P}\bigg(\bigg[\frac{1}{P_s(\theta,t_{0})}+\frac{1}{P_s(\theta,t_{0}+t_{\rm tra})}+t_{\rm tra}\bigg]<t\bigg),
\end{align}	
proof completes by taking the integration using the joint PDF of the conditional success probability separately.
\end{IEEEproof}

\section{Numerical Results}
In this section, we first validate the proposed approximation for aerial users via Monte-Carlo simulations with a large number of iterations to ensure accuracy. We then validate the analysis results of the joint distribution of conditional success probability, under different velocities of aerial and ground users, with simulations. Unless stated otherwise, we use the system parameters listed herein Table \ref{par_val}.
	\begin{table}[ht]\caption{Table of Parameters}\label{par_val}
	\centering
	\begin{center}
		\resizebox{0.8\columnwidth}{!}{
			\renewcommand{\arraystretch}{1}
			\begin{tabular}{ {c} | {c} | {c}  }
				\hline
				\hline
				\textbf{Parameter} & \textbf{Symbol} & \textbf{Simulation Value}  \\ \hline
				Density of TBSs & $\lambda$ & $1$ km$^{-2}$ \\ \hline
				UAV altitude & $h$ & 100 m\\\hline
				Environment parameters (highrise urban) & $ (a,b)$ & $(27,0.08)$ \cite{al2014optimal} \\\hline
				Environment parameters (dense urban) & $(a,b)$ & $(12,0.11)$ \cite{al2014optimal} \\\hline
				Environment parameters (urban)& $(a,b)$ & $(9.6,0.16)$ \cite{al2014optimal}  \\\hline
				Environment parameters (suburban)& $(a,b)$& $(4.88,0.43)$ \cite{al2014optimal}  \\\hline
				Transmit power, noise power & $p_t$, $\sigma^2 $ & 1, $10^{-12}$ W\\\hline
				N/LoS UAV, TBS path-loss exponent & $\alpha_{ n},\alpha_{ l},\alpha$ & $4,2.1,4$ \\\hline
				N/LoS fading parameters & $m_{ n},m_{ l}$ & $1,3$ \\\hline
				N/LoS additional loss& $\eta_{ n},\eta_{ l}$ & $-20,0$ dB \\\hline
				Handover delay& $d_m$ & $0.35$ s 
				\\\hline\hline
		\end{tabular}}
	\end{center}
\end{table}

\begin{figure}[ht]
	\centering
	\subfigure[$(a,b) = (4.88,0.43)$]{\includegraphics[width = 0.49\columnwidth]{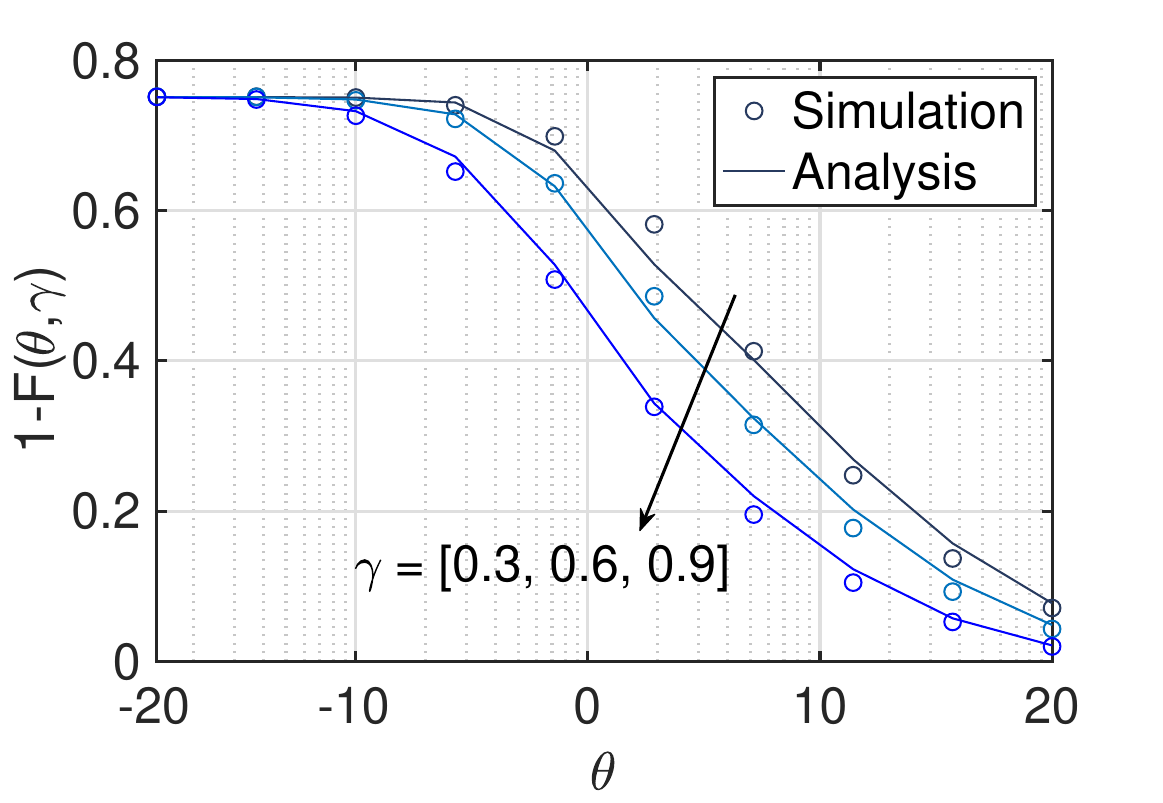}}
	\subfigure[$(a,b) = (9.6,0.16)$]{\includegraphics[width = 0.49\columnwidth]{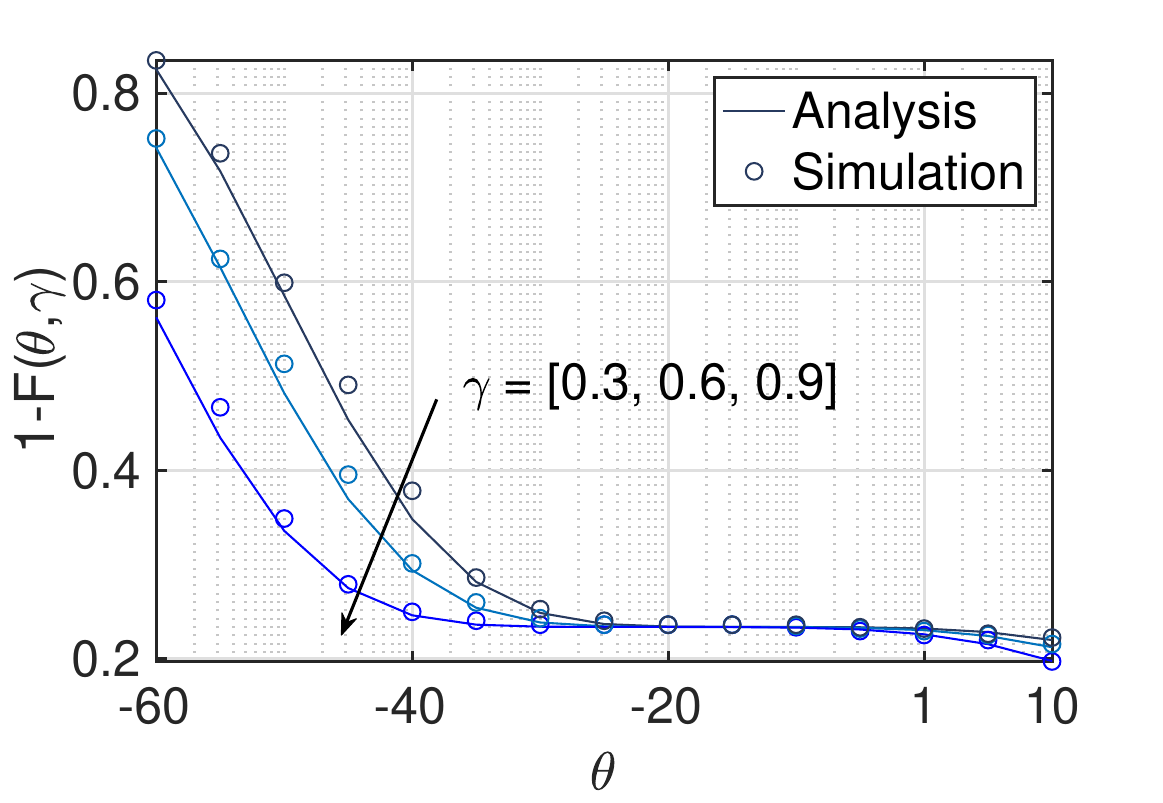}}
	\subfigure[$(a,b) = (12,0.11)$]{\includegraphics[width = 0.49\columnwidth]{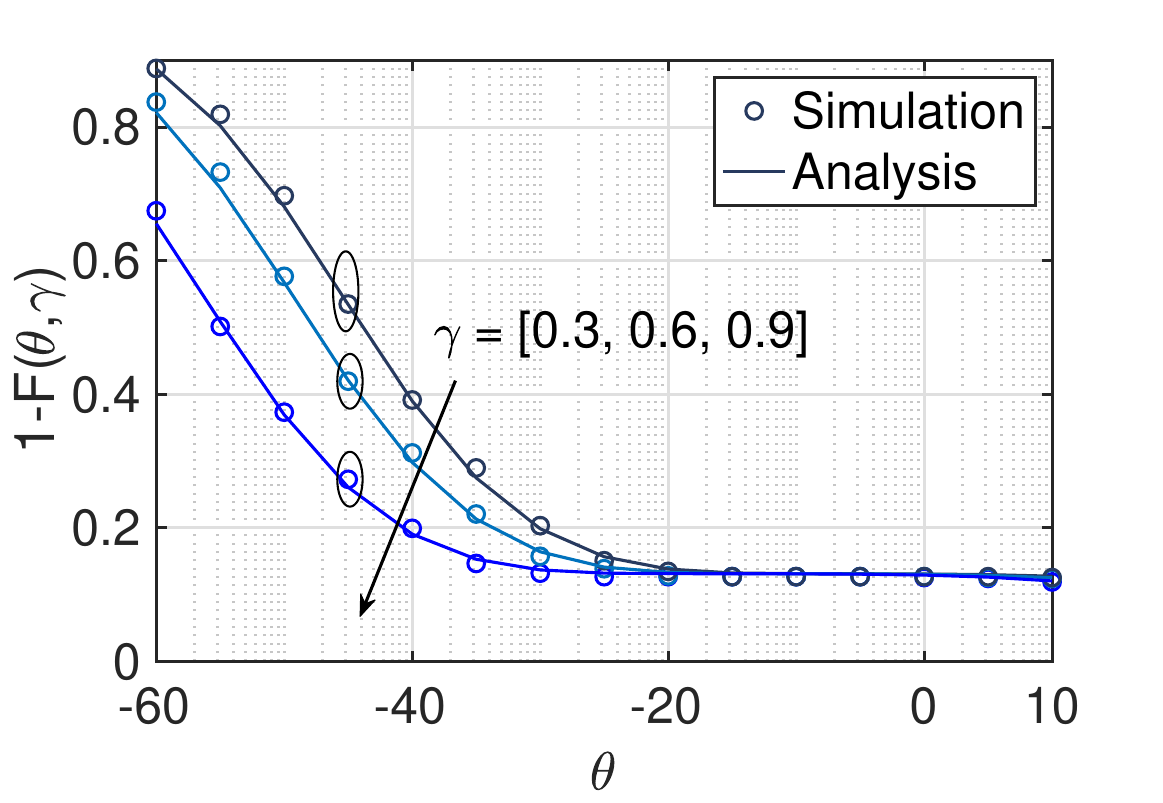}}
	\subfigure[$(a,b) = (27,0.08)$]{\includegraphics[width = 0.49\columnwidth]{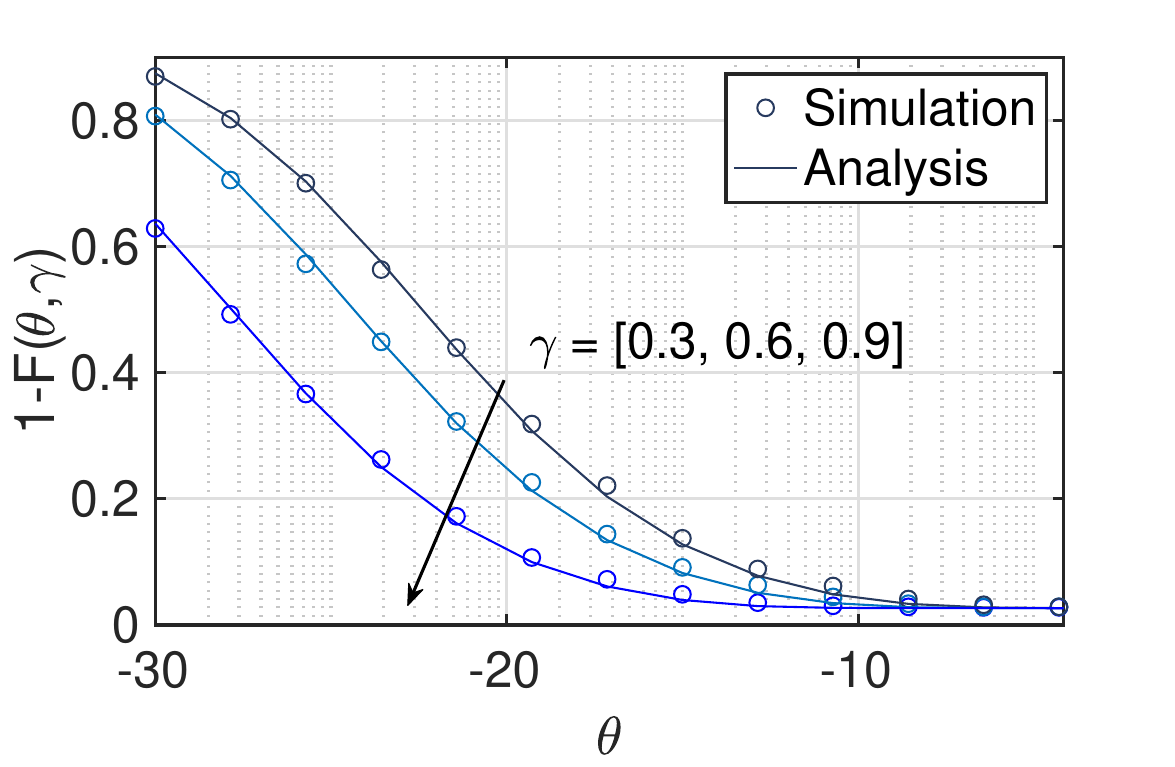}}
	\caption{Analysis results and simulation results of the SINR meta distribution of aerial users under the dominant interferer-based approximation: \textbf{(a)} suburban regions $(a,b) = (4.88,0.43)$, \textbf{(b)} urban regions $(a,b) = (9.6,0.16)$, \textbf{(c)} dense urban regions $(a,b) = (12,0.11)$, and \textbf{(a)} highrise urban regions $(a,b) = (27,0.08)$. }
	\label{fig_uav_meta_app}
	\end{figure}

In Fig. \ref{fig_uav_meta_app}, we plot the SINR meta distribution of aerial users by using the dominant interferer-based approximation under four different environments (obstacle density). By using the indicator function, the proposed approximation shows good matching for these four different environments, while we notice that the proposed approximation provides better performance in a low LoS environment compared to a high LoS environment (compare Fig. \ref{fig_uav_meta_app} (a) with Fig. \ref{fig_uav_meta_app} (d)). This is because LoS links have lower path-loss, thus higher impact of the non-dominant interferers compared to NLoS links, similar results are also shown in \cite[Fig. 4]{10066317}. In the following results, we analyze the system performance under two different environments: (i) enviro. 1, suburban regions $(a,b) = (4.88,0.43)$, and (ii) enviro. 2, highrise urban regions $(a,b) = (27,0.08)$, as they have the lowest or highest probability of establishing NLoS/LoS links, respectively.

 {\em Here, we also provide a brief summary of the dominant interferer-based approximation. While this approximation shows good matches in various scenarios (point processes), it has some limitations: (i) In the case of Rayleigh fading, the dominant interferer-based approximation can be solved using the Lambert W function \cite[(21), step (a)]{10066317}. However, in Nakagami-m fading scenarios, it can only be addressed using indicator functions, which limits the applicability of this approximation, (ii) Since this approximation is obtained by considering the nearest interferer exactly while treating other interference in an average sense, it performs better in high path-loss scenarios, , such as NLoS environments, rather than LoS scenarios. This is also noted in the above discussion, where the proposed approximation provides better performance in low LoS environments compared to high LoS environments.}

\begin{figure}[ht]
	\centering
 \includegraphics[width = 1\columnwidth]{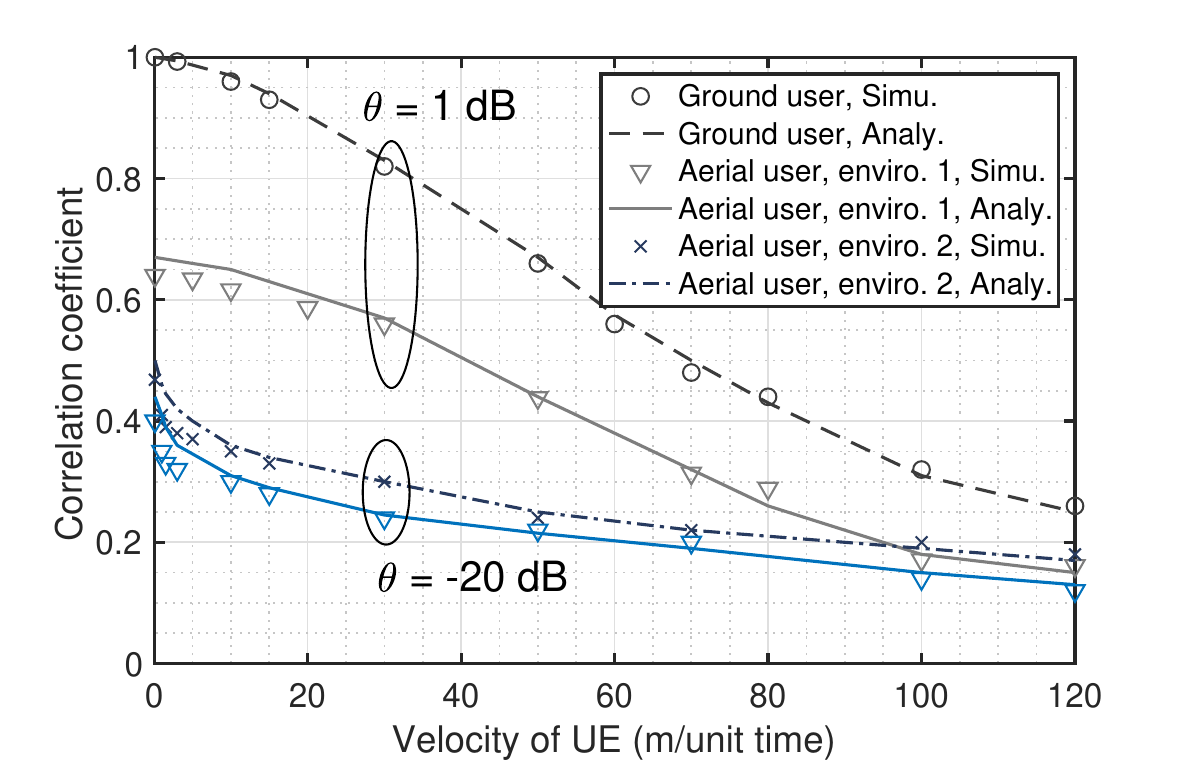}
 \caption{Analysis and simulation results of the correlation coefficient of the conditional success probability of aerial and ground users versus different velocities.}
 \label{fig_coeff}
\end{figure}
Fig. \ref{fig_coeff} depicts the spatio-temporal correlation of the conditional success probability. As the velocity of both ground and aerial users increases, the correlation of the conditional success probability decreases. This decrease is attributed to the reduced correlation among the locations of interferers. For instance, as the velocity approaches infinity, both interference and the distance to the serving BSs become totally uncorrelated. Additionally, we observe that the conditional success probability for aerial users is lower in comparison to ground users when $v = 0$ m/unit time. This is because of the probability of establishing LoS or NLoS links. In other words, although the user locations exhibit spatio-temporal correlation, they may establish different types of links. Consequently, even when the device velocity is low, the dominant interferer may be entirely different.
These results indicate that for aerial users, the network topology is more complex and dynamic compared to ground users. This underscores the importance of  analyzing and designing of dynamic interference handling techniques, as well as the importance of designing aerial UE trajectories to minimize the impact of rapid interference and signal changes due to LoS and NLoS links. Ultimately, these analysis are crucial for achieving reliable and high quality of service communications to aerial UE during implementation.

\begin{figure}[ht]
	\centering
\includegraphics[width = 1\columnwidth]{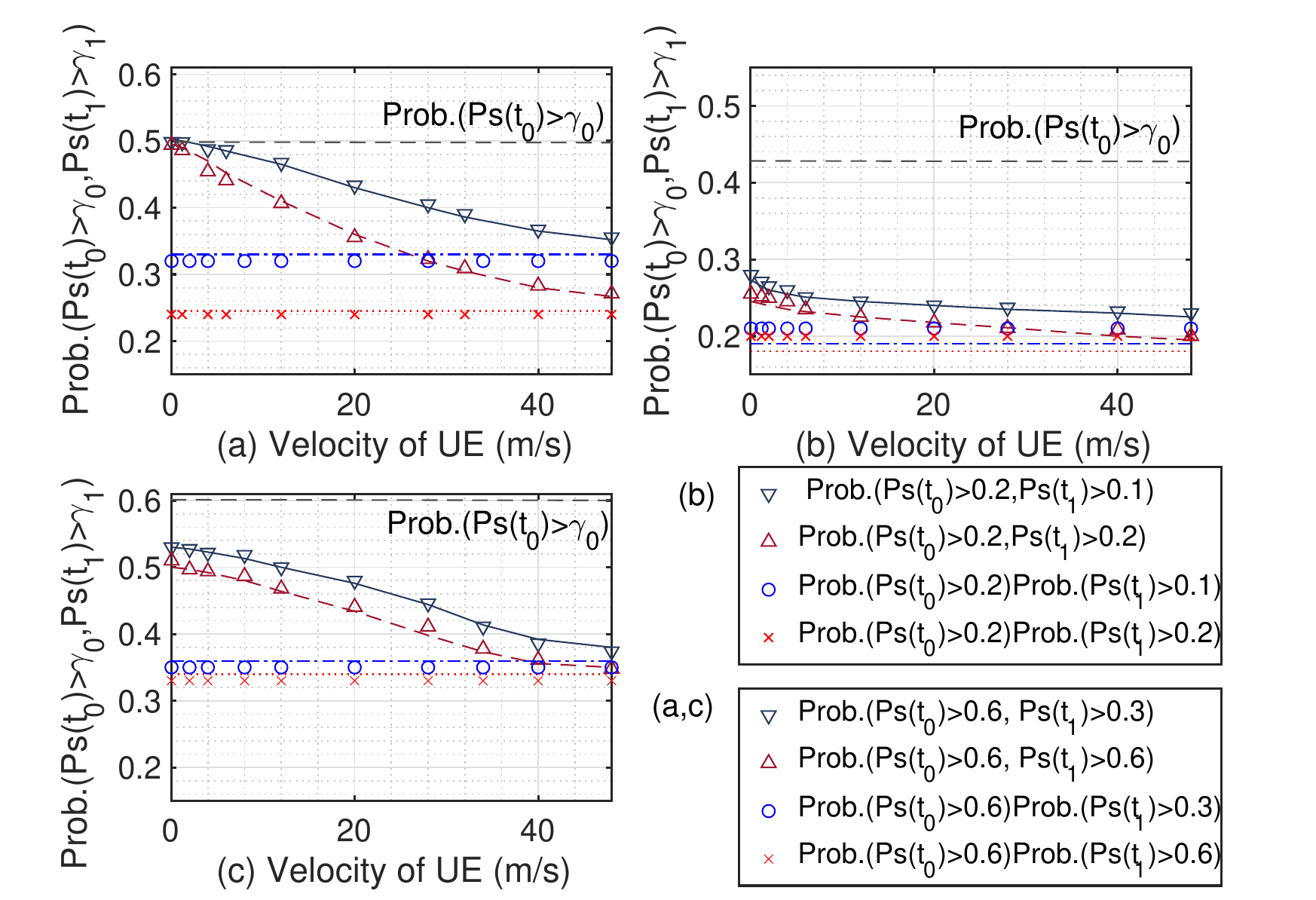}
 \caption{Analysis and simulation results of the joint distribution of the conditional success probability of \textbf{(a)} ground users, \textbf{(b)} aerial users in highrise urban regions $(a,b) = (27,0.08)$, and \textbf{(c)} aerial users in suburban regions $(a,b) = (4.88,0.43)$.}
\label{fig_jointmetadis}
\end{figure}
We then illustrate the impact of user velocity on the joint distribution of the conditional success probability in Fig. \ref{fig_jointmetadis}. Fig. \ref{fig_jointmetadis} (a) and Fig. \ref{fig_jointmetadis} (c) are plotted with a SINR threshold of $\theta = 0$ dB, while Fig. \ref{fig_jointmetadis} (b) is plotted with a SINR threshold of $\theta = -20$ dB \footnote{The reason for using $\theta = -20$ dB for highrise urban regions is that the conditional success probability is too low when $\theta = 0$ dB , as shown in Fig. \ref{fig_uav_meta_app} (d).}. As the velocity increases, the joint probability of the conditional success probability decreases rapidly at first and then more slowly. This behavior is due to the correlation decreasing as velocity increases. For example, at high velocities, $\mathbb{P}(P_s(t_0) > \gamma_0, P_s(t_1) > \gamma_1)$ is approximately equal to $\mathbb{P}(P_s(t_0) > \gamma_0) \times \mathbb{P}(P_s(t_1) > \gamma_1)$. Furthermore, we observe that aerial users in highrise urban regions with parameters $(a, b) = (27, 0.08)$ exhibit the worst performance, even when compared to ground users. This is because in such high-dense populated or obstacle-rich environments, the probability of establishing LoS links is very low (e.g., $P_l = 0.02$, obtained by averaging over the distance to the serving BS). However, the power of interference increases significantly in this scenario.

\begin{figure}[ht]
	\centering
	\includegraphics[width = 1\columnwidth]{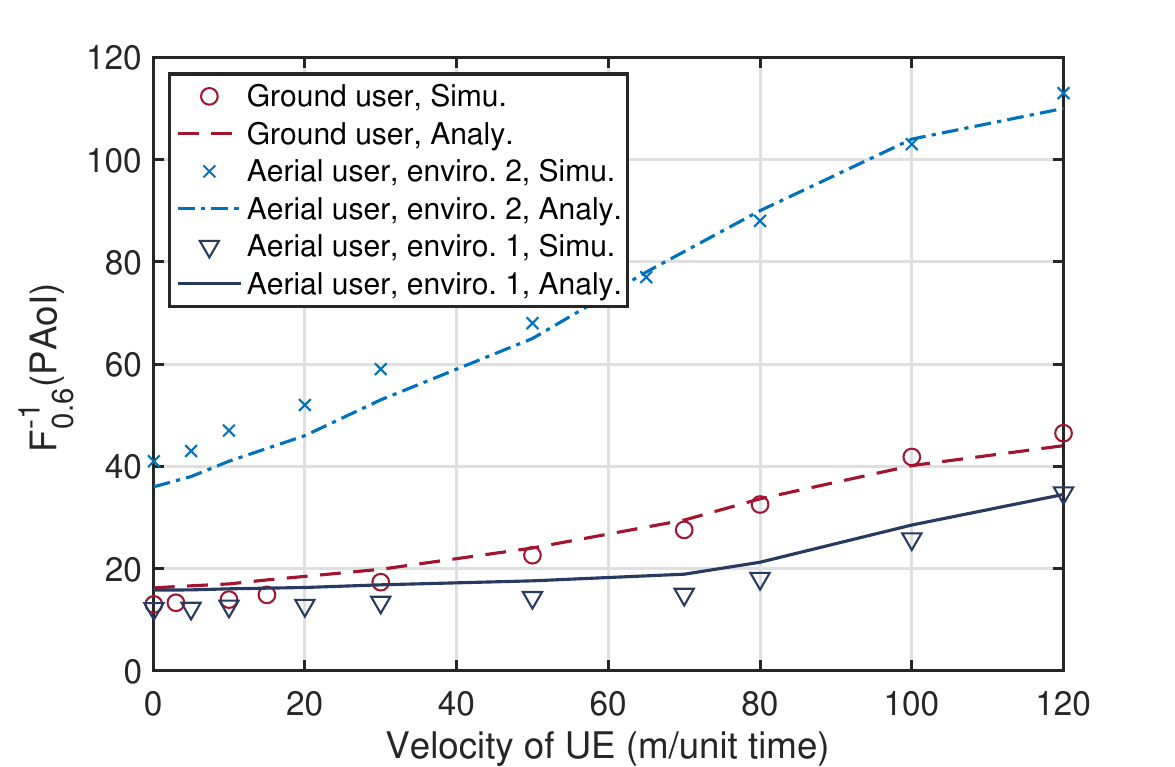}
	 \caption{Analysis and simulation results of the $60$-th-percentile of PAoI of aerial and ground users versus different velocities.}
	\label{fig_CDF_paoi}
\end{figure}
In Fig. \ref{fig_CDF_paoi}, we show the impact of velocity on the $60$-th-percentile of PAoI. For aerial users in enviro. 2, the SINR threshold is set at $\theta = -20$ dB, and, for aerial users in enviro. 1 and ground users, the SINR threshold is set at $\theta = 0$ dB. As expected, the $60$-th-percentile of PAoI increases with the increase of the velocities due to the decrease in spatio-temporal correlation and the decrease in the joint probability of conditional success probability. We also observe that the $60$-th-percentile of PAoI increases gradually at both low and high velocity values, which is similar to the decreasing trends in the joint probability of conditional success probability and correlation coefficient. Additionally, because the spatio-temporal correlation of aerial users in highrise urban regions $(a, b) = (27, 0.08)$ is comparatively low and decreases slowly, the $60$-th-percentile  of PAoI for aerial users in highrise urban regions is less sensitive compared to the other two types of users.
 These results highlight that velocity has a significant impact when collecting monitoring data from aerial IoT UEs. Designing redundant systems, such as using multiple devices to monitor a single physical process, can help minimize PAoI. Additionally, optimizing UE trajectory design based on conditional success probability, such as flying within association zones to maximize minimal SINR or success probability (as discussed in \cite{8531711,10225703}), can also effectively reduce PAoI.

\section{Conclusion}

This study introduces a framework to analyze the impact of spatio-temporal correlation, specifically user velocity, on the conditional success probability for both ground and aerial users using the dominant interferer-based approximation. Initially, we estimate the SINR meta distribution based on distances to the serving base station and the dominant interferer. Our results reveal higher spatio-temporal correlations among ground users, emphasizing the  impact of velocity on  the joint probability of conditional success and PAoI distribution. Furthermore, our work demonstrates the strong matching performance of the dominant interferer-based approximation in complex scenarios, such as the Nakagami-m fading model, making it effective in computing spatio-temporal correlation with low complexity. In summary, this approximation holds significant potential for accurately analyzing network performance under varying conditions.

{\color{blue}
	This work has many possible extensions. Firstly, the resulting expressions are not in closed form and require numerical integration. It would be valuable to extend this work by deriving easy-to-use approximations and bounds, which would allow readers to gain even better insights. Secondly, while we have shown that velocity impacts the distribution of the joint conditional success probability, and have used PAoI as the performance metric to demonstrate this effect, other performance metrics, such as error percentages, should also be analyzed. Additionally, our primary focus is on analyzing PAoI under the basic handover strategy, where handovers occur when the nearest BS changes (referred to as the best connected strategy in \cite{7827020}). However, other handover strategies, such as those based on biased received power or skipping certain cells, which have shown lower handover costs compared to the basic strategy, could be investigated in future research. In addition to PAoI under different handover strategies, exploring PAoI under various interference handling techniques, such as interference cancellation, is also a potential area of analysis. Furthermore, investigating PAoI under optimal UE trajectory planning, considering interference minimization \cite{wu2018joint} instead of a straight-line trajectory as assumed in this work, presents another intriguing avenue for future investigation. Moreover, dynamic velocities and changing moving directions can also be an interesting future topic.
}
\appendix
\subsection{Proof of Theorem \ref{theo_JointPs_GU}} \label{app_JointPs_GU}
The joint probability $\mathbb{P}(P_s(\theta,t_0)>\gamma_0,P_s(\theta,t_1)>\gamma_1)$ can be represented by using the conditional probability as follows:
\begin{align}
	&\mathbb{P}(P_s(\theta,t_0)>\gamma_0,P_s(\theta,t_1)>\gamma_1) \nonumber\\
	&= \mathbb{P}(P_s(\theta,t_1)>\gamma_1 \mid P_s(\theta,t_0)>\gamma_0 )\mathbb{P}(P_s(\theta,t_0)>\gamma_0),
\end{align}
now our goal is to compute the conditional probability and $\mathbb{P}(P_s(\theta,t_0)>\gamma_0)$.

We first notice that if we use the dominant interferer-based approximation derived in \cite[Eq. (22)]{10066317}, at time $t_0$, the approximated conditional success probability is a function of the distance to the serving BS $R_{0}(t_0)$ and the distance to the first nearest interferer $R_{1}(t_0)$. With that being said, as for a given $\gamma_0$, the SINR meta distribution, $\mathbb{P}(P_s(\theta,t_0)>\gamma_0)$, equals to the probability of $R_{0}(t_0)$ and $R_{1}(t_0)$ satisfy a certain condition:
\begin{align}
 & \mathbb{P}(P_s(\theta,t_0)>\gamma_0)= \mathbb{P}(R_{0}(t_0)<K(R_{1}(t_0),\gamma_0,\theta))\label{1}\\
&= \int_{0}^{\infty} F_{R_0(t_0)}(K(r_1,\gamma_0,\theta))  f_{R_1(t_0)}(r_1){\rm d}r_1,\label{eq_gu_no_2}
\end{align}
where
\begin{align}
&K(R_{1}(t_0),\gamma_0,\theta) = \Bigg(-\frac{1}{\theta R_{1}^{-\alpha}(t_0)}+\frac{1}{s(R_{1}(t_0))}\nonumber\\
&\times W\bigg(0,\frac{s(R_{1}(t_0))\exp(s(R_{1}(t_0))\frac{R_{1}^{\alpha}(t_0)}{\theta})}{\gamma_0\theta R_{1}^{-\alpha}(t_0) }\bigg)\Bigg)^{\frac{1}{\alpha}},
\end{align}
similar to time $t_1$. Therefore, the joint distribution now becomes
\begin{align}
	\label{eq_jointPs_proof1}
&\mathbb{P}(P_s(\theta,t_1)>\gamma_1\mid P_s(\theta,t_0)>\gamma_0 )\nonumber\\
 &= \mathbb{P}(R_{0}(t_1)<K(R_{1}(t_1),\gamma_1,\theta)\mid R_{0}(t_0)<K(R_{1}(t_0),\gamma_0,\theta)).
\end{align}
In the case of no handover happens, we substitude (\ref{1}) into  (\ref{eq_jointPs_proof1}),
\begin{align}
&\mathbb{P}(R_{0}(t_1)<K(R_{1}(t_1),\gamma_1,\theta)\mid R_{0}(t_0)<K(R_{1}(t_0),\gamma_0,\theta)) \nonumber\\
&=\mathbb{E}\bigg[F_{R_0(t_1)|R_0(t_0)}(K(R_1(t_1),\gamma_1,\theta))\bigg]\nonumber\\
&=\mathbb{E}\bigg[F_{R_0(t_1)|R_0(t_0)}(K^{\prime}(R_1(t_0),\kappa_1,\gamma_1,\theta))\bigg]\nonumber\\
&=\mathbb{E}_{\{R_0(t_0),R_1(t_0),\kappa_1\}}\bigg[\frac{\kappa(R_0(t_0),K^{\prime}(R_1(t_0),\kappa_1,\gamma_1,\theta))}{\pi}\bigg]\nonumber\\
&=\int_{0}^{\infty}\int_{0}^{K(r_1,\gamma_0,\theta)}\int_{0}^{\pi}\frac{1}{\pi}\frac{\kappa(r_0,K^{\prime}(r_1,\kappa_1,\gamma_1,\theta))}{\pi}{\rm d}\kappa_1\nonumber\\
&\quad \times f_{R_0(t_0)}^{\prime}(r_0){\rm d}r_0 f_{R_1(t_0)}(r_1){\rm d}r_1,\label{eq_gu_no_1}
\end{align}
in which $K^{\prime}(R_1(t_0),\kappa_1,\gamma_1,\theta) = K(g^{-1}(R_1(t_0)),\gamma_1,\theta)= K(R_1(t_1),\gamma_1,\theta)$, where $g(\cdot)$ is the function defined in (\ref{eq_R0R1}), and the truncated distribution of $R_{0}(t_0)$ is 
\begin{align}
	f_{R_{0}(t_0)}^{\prime}(r) &= \frac{f_{R_{0}(t_0)}(r)}{F_{R_{0}(t_0)}(0)-F_{R_{0}(t_0)}(K(R_{1}(t_0),\gamma_0,\theta))} \nonumber\\
	&= \frac{f_{R_{0}(t_0)}(r)}{\exp(-\pi\lambda K^{2}(R_{1}(t_0),\gamma_0,\theta) )}.
\end{align}
Consequently, the joint distribution is obtained by the multiplication of (\ref{eq_gu_no_2}) and (\ref{eq_gu_no_1}).

When the handover happens, we assume that the interference is uncorrelated due to the resource allocated randomly,
\begin{align}
	&\mathbb{P}(P_s(\theta,t_1)>\gamma_1\mid P_s(\theta,t_0)>\gamma_0 )\nonumber\\
	& \approx \mathbb{P}(R_{0}^{\prime}(t_1)<K(R_{1}^{\prime}(t_1),\gamma_1,\theta)\mid R_{0}(t_0)<K(R_{1}(t_0),\gamma_0,\theta))\nonumber\\
	&= \int_{0}^{\infty}\int_{0}^{\infty}\int_{0}^{\pi}\int_{0}^{K(z_0,\gamma_0,\theta)} F_{R_{0}^{\prime}|\kappa_0,R_0}(K(z_1,\gamma_1,\theta),\kappa,r_0)\nonumber\\
	&\quad\times\frac{1}{\pi}{\rm d}\kappa f_{R_1(t_0)}(z_1){\rm d}z_1 f_{R_0(t_0)}(r_0){\rm d}r_0 f_{R_1(t_0)}(z_0){\rm d}z_0,\label{eq_gu_h_1}
\end{align}
where
\begin{align}
	&R_{d} = \sqrt{(vt)^{2}+R_{0}^{2}(t_0) -2vtR_{0}(t_0) \cos(\kappa_0)},\nonumber\\
	&f_{R_{0}^{\prime}|\kappa_0,R_0}(r,\kappa_0,R_0) =  \frac{2\pi\lambda r\exp(-\pi\lambda r^2)}{\exp(-\lambda\pi R_{d}^{ 2})}, r<R_{d}^{ 2},\nonumber\\
	&F_{R_{0}^{\prime}|\kappa_0,R_0}(r,\kappa_0,R_0) =  \frac{1-\exp(-\pi\lambda r^2)}{\exp(-\lambda\pi R_{d}^{ 2})}, r<R_{d}^{ 2}.\label{eq_57}
\end{align}
Consequently, the joint distribution is obtained by the multiplication of (\ref{eq_gu_no_2}) and (\ref{eq_gu_h_1}).

\subsection{Proof of Lemma \ref{lemma_condPs_au}}\label{app_condPs_au}
By using the dominant interferer-based approximation, the conditional success probability of the aerial UE is approximated by
\begin{align}
	&P_{s,ll}(\theta) = \mathbb{P}\bigg(\frac{p_t \eta_l G_l (R_{0}^2+h^2)^{-\frac{\alpha_l}{2}}}{I+\sigma^2}>\theta\mid \Phi_i\bigg)\nonumber\\
	&= \mathbb{E}_{I}\bigg[\frac{\Gamma_{u}(m_l,m_l \theta(I+\sigma^2)(p_t \eta_l (R_{0}^2+h^2)^{-\frac{\alpha_l}{2}})^{-1})}{\Gamma(m_l)}\bigg]\nonumber\\
	&\stackrel{(a)}{\approx} \mathbb{E}_{G_l}\left[\frac{\Gamma_{u}\left(m_l,\frac{m_l \theta(p_t \eta_{ l} G_l (R_{l}^2+h^2)^{-\frac{\alpha_{l}}{2}}+I_{l}(R_l)+\sigma^2)}{p_t \eta_l (R_{0}^2+h^2)^{-\frac{\alpha_l}{2}}}\right)}{\Gamma(m_l)}\right]\nonumber\\
	&\stackrel{(b)}{\approx} \mathbb{E}_{G_l}\bigg[\sum_{k = 1}^{m_l}\binom{m_l}{k}(-1)^{k+1}\exp\bigg(-k\beta_l \nonumber\\
	&\quad\times\frac{m_l \theta(p_t \eta_{ l} G_l (R_{l}^2+h^2)^{-\frac{\alpha_{l}}{2}}+I_{l}(R_l)+\sigma^2)}{p_t \eta_l (R_{0}^2+h^2)^{-\frac{\alpha_l}{2}}}\bigg)\bigg]\nonumber\\
	&= \sum_{k = 1}^{m_l}\binom{m_l}{k}(-1)^{k+1}\exp\bigg(-\frac{k\beta_l m_l \theta(I_{l}(R_l)+\sigma^2)}{p_t \eta_l (R_{0}^2+h^2)^{-\frac{\alpha_l}{2}}}\bigg)\nonumber\\
	&\quad\times\mathbb{E}_{G_l}\bigg[\exp\bigg(-k\beta_l m_l \theta G_l (R_{l}^2+h^2)^{-\frac{\alpha_{l}}{2}} (R_{0}^2+h^2)^{\frac{\alpha_l}{2}}\bigg)\bigg]\nonumber\\
	&= \sum_{k = 1}^{m_l}\binom{m_l}{k}(-1)^{k+1}\exp\bigg(-\frac{k\beta_l m_l \theta(I_{l}(R_l)+\sigma^2)}{p_t \eta_l (R_{0}^2+h^2)^{-\frac{\alpha_l}{2}}}\bigg)\nonumber\\
	&\quad\times(1+k\beta_l\theta (R_{l}^2+h^2)^{-\frac{\alpha_{l}}{2}}(R_{0}^2+h^2)^{\frac{\alpha_l}{2}})^{-m_l},
\end{align}
where step (a) follows from using the dominant interferer-based approximation, and step (b) follows from using the upper bound of the CDF of Gamma distribution and binomial theorem,
similar to $P_{s,ln}(\theta)$, which is derived by
\begin{align}
&P_{s,ln}(\theta) = \mathbb{P}\bigg(\frac{p_t \eta_l G_l (R_{0}^2+h^2)^{-\frac{\alpha_l}{2}}}{I+\sigma^2}>\theta\mid \Phi_i\bigg)\nonumber\\
&\approx \mathbb{E}_{I}\left[\frac{\Gamma_{u}\left(m_l,\frac{m_l \theta(p_t \eta_{ n} G_n (R_{n}^2+h^2)^{-\frac{\alpha_{n}}{2}}+I_{n}(R_n)+\sigma^2)}{p_t \eta_l (R_{0}^2+h^2)^{-\frac{\alpha_l}{2}}}\right)}{\Gamma(m_l)}\right]\nonumber\\
&\approx \mathbb{E}_{G_n}\bigg[\sum_{k = 1}^{m_l}\binom{m_l}{k}(-1)^{k+1}\exp\bigg(-k\beta_l \nonumber\\
&\quad\times\frac{m_l \theta(p_t \eta_{ n} G_n (R_{n}^2+h^2)^{-\frac{\alpha_{n}}{2}}+I_{n}(R_n)+\sigma^2)}{p_t \eta_l (R_{0}^2+h^2)^{-\frac{\alpha_l}{2}}}\bigg)\bigg]\nonumber\\
&= \sum_{k = 1}^{m_l}\binom{m_l}{k}(-1)^{k+1}\exp\bigg(-\frac{k\beta_l m_l \theta(I_{l}(R_1)+\sigma^2)}{p_t \eta_l R_{u}^{-\alpha_{l}}}\bigg)\nonumber\\
&\mathbb{E}_{G_n}\bigg[\exp\bigg(-k\beta_l m_l \theta G_n\eta_n (R_{n}^2+h^2)^{-\frac{\alpha_{n}}{2}} (R_{0}^2+h^2)^{\frac{\alpha_l}{2}}\bigg)\bigg]\nonumber\\
&= \sum_{k = 1}^{m_l}\binom{m_l}{k}(-1)^{k+1}\exp\bigg(-\frac{k\beta_l m_l \theta(I_{n}(R_n)+\sigma^2)}{p_t \eta_l (R_{0}^2+h^2)^{-\frac{\alpha_l}{2}}}\bigg)\nonumber\\
&\quad\times(1+k\beta_l\theta m_l (R_{n}^2+h^2)^{-\frac{\alpha_{n}}{2}}(R_{0}^2+h^2)^{\frac{\alpha_l}{2}}\eta_n)^{-m_n},
\end{align}
and $P_{s,nl}(\theta)$ and $P_{s,nn}(\theta)$ can be obtained by similar steps or setting $m_l = m_n$.
\subsection{Proof of Theorem \ref{theorem_JointPs_au}}
\label{app_JointPs_au}
The joint distribution of aerial UE is obtained by,
\begin{small}
\begin{align}
	\label{eq_proof_jointPs_au}
	&\mathbb{P}(P_s(\theta,t_1)>\gamma_1, P_s(\theta,t_0)>\gamma_0 ) \nonumber\\
	&\approx \mathbb{P}([P_{s,ll}(\theta,t_1)+P_{s,ln}(\theta,t_1)+P_{s,nl}(\theta,t_1)+P_{s,nn}(\theta,t_1)]>\gamma_1, \nonumber\\
	&\quad [P_{s,ll}(\theta,t_0)+P_{s,ln}(\theta,t_0)+P_{s,nl}(\theta,t_0)+P_{s,nn}(\theta,t_0)]>\gamma_0)\nonumber\\
	& = \sum_{i_0,j_0,i_1,j_1\in\{l,n\}}\mathbb{E}[\mathcal{A}_{i_0j_0}(R_0(t_0),R_{j_0})\mathcal{A}_{i_1j_1}(R_0(t_1),R_{j_1})\nonumber\\
	&\qquad\times \mathbb{P}( \kappa_{i_0j_0}(R_0(t_0),R_{j_0})>\gamma_0, \kappa_{i_1j_1}(R_0(t_1),R_{j_1})>\gamma_1)],
\end{align}
\end{small}
in the case of no handover, (\ref{eq_proof_jointPs_au}) is computed by
\begin{small}
\begin{align}
	 &=\sum_{i_0,j_0,i_1,j_1\in\{l,n\}}\mathbb{E}[\mathcal{A}_{i_0j_0}(R_0(t_0),R_{j_0})\mathcal{A}_{i_1j_1}(g(R_0(t_0),\kappa_0),R_{j_1})\nonumber\\
	&\times \mathbb{P}( \kappa_{i_0j_0}(R_0(t_0),R_{j_0})>\gamma_0, \kappa_{i_1j_1}(g(R_0(t_0),\kappa_0),R_{j_1})>\gamma_1)]\nonumber\\
	& = \sum_{i_0,j_0,i_1,j_1\in\{l,n\}}\int_{0}^{\pi}\iiint_{\mathbbm{1}(c_{h}^{\prime})}\mathcal{A}_{i_0j_0}(r_0,z_0)\mathcal{A}_{i_1j_1}(g(r_0,\kappa),z_1)\nonumber\\
	&\times \frac{1}{\pi}f_{R_0(t_0)}(r_0){\rm d}r_0 f_{R_{j_0}}(z_0){\rm d}z_0 f_{R_{j_1}}(z_1){\rm d}z_1{\rm d}\kappa,
\end{align}
\end{small}
where $c_{h}^{\prime} = \kappa_{i_0j_0}(r_0,z_0)>\gamma_0, \kappa_{i_1j_1}(g(r_0,\kappa_0),z_1)>\gamma_1$,
in the case of handover happens, (\ref{eq_proof_jointPs_au}) is computed by
\begin{align}
	&=\sum_{i_0,j_0,i_1,j_1\in\{l,n\}}\mathbb{E}[\mathcal{A}_{i_0j_0}(R_0(t_0),R_{j_0})\mathcal{A}_{i_1j_1}(R_{0}^{\prime}(t_1),R_{j_1})\nonumber\\
	&\times \mathbb{P}( \kappa_{i_0j_0}(R_0(t_0),R_{j_0})>\gamma_0, \kappa_{i_1j_1}(R_{0}^{\prime}(t_1),R_{j_1})>\gamma_1)]\nonumber\\
	& \approx \sum_{i_0,j_0,i_1,j_1\in\{l,n\}}\iiiint_{\mathbbm{1}(c_{h})}\mathcal{A}_{i_0j_0}(r_0,z_0)\mathcal{A}_{i_1j_1}(r_1,z_1)\nonumber\\
	&\times f_{R_1|\kappa_0,R_0}(r_1,\pi,r_0)f_{R_0(t_0)}(r_1){\rm d}r_1 f_{R_0(t_0)}(r_0){\rm d}r_0\nonumber\\
	&\times f_{R_{j_0}}(z_0){\rm d}z_0 f_{R_{j_1}}(z_1){\rm d}z_1,
\end{align}
where $c_{h} = \kappa_{i_0j_0}(r_0,z_0)>\gamma_0, \kappa_{i_1j_1}(r_1,z_1)>\gamma_1$, and the approximation sign follows from we set $\kappa_0 = \pi$ to reduce the number of integrations.

\bibliographystyle{IEEEtran}
\bibliography{Rep19}
\end{document}

%% file: notation.tex
\def\nb0{{\mathbf{0}}}
\def\nb1{{\mathbf{1}}}

\newtheorem{lemma}{Lemma}

\newtheorem{definition}{Definition}

\newtheorem{theorem}{Theorem}

\newtheorem{remark}{Remark}
\newtheorem{assumption}{Assumption}
	
